%% file: main.tex
\documentclass[fleqn,10pt]{wlscirep}
\usepackage[utf8]{inputenc}
\usepackage[T1]{fontenc}
\usepackage{arydshln}
\usepackage[version=4]{mhchem}

\usepackage{color}

\title{Accelerated Inorganic Materials Design with Generative AI Agents}

\author[1,2,3,*]{Izumi Takahara}
\author[1]{Teruyasu Mizoguchi}
\author[2,3]{Bang Liu}
\affil[1]{The University of Tokyo, Institute of Industrial Science, Tokyo, 153-8505, Japan}
\affil[2]{University of Montreal, Department of Computer Science and Operations Research, Montreal, H3T 1J4, Canada}
\affil[3]{Mila - Quebec AI Institute, Montreal, H2S 3H1, Canada}

\affil[*]{kougen@iis.u-tokyo.ac.jp}

\newcommand{\framework}{MatAgent }
\newcommand{\frameworkws}{MatAgent}

\begin{abstract}
Designing inorganic crystalline materials with tailored properties is critical to technological innovation, yet current generative computational methods often struggle to efficiently explore desired targets with sufficient interpretability. Here, we present \frameworkws, a generative approach for inorganic materials discovery that harnesses the powerful reasoning capabilities of large language models (LLMs). By combining a diffusion-based generative model for crystal structure estimation with a predictive model for property evaluation, \framework uses iterative, feedback-driven guidance to steer material exploration precisely toward user-defined targets. Integrated with external cognitive tools—including short-term memory, long-term memory, the periodic table, and a comprehensive materials knowledge base—\framework emulates human expert reasoning to vastly expand the accessible compositional space. Our results demonstrate that \framework robustly directs exploration toward desired properties while consistently achieving high compositional validity, uniqueness, and material novelty. This framework thus provides a highly interpretable, practical, and versatile AI-driven solution to accelerate the discovery and design of next-generation inorganic materials.
\end{abstract}
\begin{document}

\flushbottom
\maketitle
%
%
\thispagestyle{empty}

\input{body/0_introduction}
\input{body/1_results}
\input{body/2_discussion}
\input{body/3_methods}

\bibliography{main}

\section*{Acknowledgements}
The authors thank Qianggang Ding and Huan Zhang for helpful discussions. This work was supported by JST ACT-X (Grant Number JPMJAX24DB) and JST BOOST (Grant Number JPMJBS2418), Japan. I. T. was supported by the MERIT-WINGS, The University of Tokyo.

\section*{Author contributions statement}
I. T. and B. L. designed the study. I. T. developed the framework, conducted the experiments, analyzed results, and conducted manuscript drafting. T. M. assisted the framework evaluation and contributed to the manuscript editing. B. L. supervised the study and contributed to the manuscript editing. All authors reviewed the manuscript.

\section*{Code availability}
The source code and the dataset used in this study are publicly available on
GitHub (https://github.com/izumitkhr/matagent).

\section*{Additional information}

\textbf{Competing interests} The authors have no conflicts to disclose.

\newpage
\appendix
\input{body/4_appendix_1}
\input{body/4_appendix_3}
\input{body/4_appendix_2}

\end{document}

%% file: body/0_introduction.tex
\section*{Introduction}

Developing materials with desired properties is a critical challenge, driving innovations across essential industrial fields such as catalysis, energy storage, and beyond\cite{chen2020critical, toyao2020machine, Merchant2023scaling}. Computational approaches, including density functional theory\cite{hohenberg1964inhomogeneous,kohn1965self} simulations and high-throughput virtual screenings, have been widely employed to identify candidate materials\cite{noh2019machine}. However, these traditional methods typically require extensive manual modeling by human experts or exhaustive enumeration of candidate materials, limiting their efficiency and scalability in exploring broader materials spaces. To overcome these limitations and facilitate more autonomous, scalable, and efficient materials discovery, there is an increasing demand for workflows capable of effectively exploring broader materials space to identify novel inorganic materials while reducing the reliance on human-expert intervention.

Recently, generative artificial intelligence has made significant progress, particularly in the fields of natural language processing and computer vision. These advances have also influenced computational materials design, where generative models such as variational autoencoders\cite{kingma2014auto}, generative adversarial networks\cite{goodfellow2014generative}, generative flow networks\cite{bengio2021flow}, diffusion models\cite{sohldickstein2015deep,ho2020ddpm, song2021score}, and autoregressive models\cite{vaswani2017attention} are increasingly being applied, individually\cite{Ren2022inv, kim2020generative, ai4science2023crystalgfn, yang2024scalable, gruver2024fine, Antunes2024, miller2024flowmm, park2025multi, zhao2021high, zhao2023physics, zhu2024wycryst, Zeni2025} or in combination\cite{xie2022crystal, luo2023symat, sriram2024flow, luo2024deep}, to generate and explore novel crystalline materials. This opens new avenues for effectively navigating and exploring the vast materials design space utilizing generative artificial intelligence techniques\cite{park2024exploration,yang2024gen}. For example, MatterGen\cite{Zeni2025}, a diffusion-based generative model, has successfully generated crystal structures by simultaneously generating lattice vectors, atomic coordinates, and elemental species, enabling the exploration and discovery of novel materials beyond conventional approaches. Similarly, large language models (LLMs), fine-tuned by Gruver \textit{et al.}\cite{gruver2024fine} or trained from scratch by Antunes \textit{et al.}\cite{Antunes2024}, have directly generated stable materials, highlighting the broad potential of generative AI approaches in accelerating materials discovery.

However, these generative approaches typically produce materials through a single-step generation process, potentially limiting their ability to precisely meet specified target properties. Furthermore, in many target-aware generation methods, materials are optimized within latent spaces, making it challenging to interpret the underlying reasoning behind the generated results. Recent Advances in the foundational knowledge and reasoning capabilities of LLMs have enabled these models to perform complex and sophisticated tasks\cite{openai2024gpt4, shinn2023reflexion, yao2023react}. Leveraging these enhanced capabilities, multi-step refinement approaches have emerged as a promising direction. For instance, MatExpert\cite{ding2024matexpert} decomposed the material generation process into three distinct steps and demonstrated improved performance in generating target-oriented materials. Jia \textit{et al.} proposed an LLM-driven iterative framework that refines materials via modifications, accompanied by human-interpretable reasoning provided in natural language\cite{jia2024llmatdesign}. Nevertheless, existing LLM-based methods primarily rely on the inherent knowledge embedded within the LLMs, thereby constraining their exploration to relatively limited compositional spaces or frequently requiring fine-tuning tailored specifically to each target-specific property or constraint, thus reducing their practical scalability. To achieve more effective materials discovery, it is essential to develop scalable frameworks capable of autonomously exploring broader materials spaces, while incorporating external knowledge and enabling interpretable, iterative refinement toward target properties.

In this study, we propose a framework named \frameworkws, which employs a powerful and general-purpose LLM as the central generative engine. By leveraging the LLM's capability for explicit reasoning in composition proposals, our framework ensures interpretability in the material design process. Inspired by the reasoning process of human experts, the framework is further enhanced with four external tools—--including short-term memory, long-term memory, the periodic table, and a materials knowledge base—--enabling exploration beyond the LLM's inherent knowledge and into a broader materials space. In addition, by integrating a generative model for crystal structure estimation and a predictive model for property evaluation, the framework enables feedback-driven refinement of proposed compositions, allowing for autonomous exploration in the materials design space. We demonstrate the effectiveness of this framework in the task of proposing materials with target properties. Furthermore, we show that the framework can be seamlessly integrated with natural language, making it suitable for practical and efficient materials design workflows.

%% file: body/1_results.tex
\section*{Results}

\input{body/1_results_1_framework}
\input{body/1_results_2_evaluation}
\input{body/1_results_3_init}
\input{body/1_results_4_constraint}





 




%% file: body/1_results_1_framework.tex
\subsection*{\framework  framework}
\begin{figure}[htbp]
    \centering
    \includegraphics[width=\linewidth]{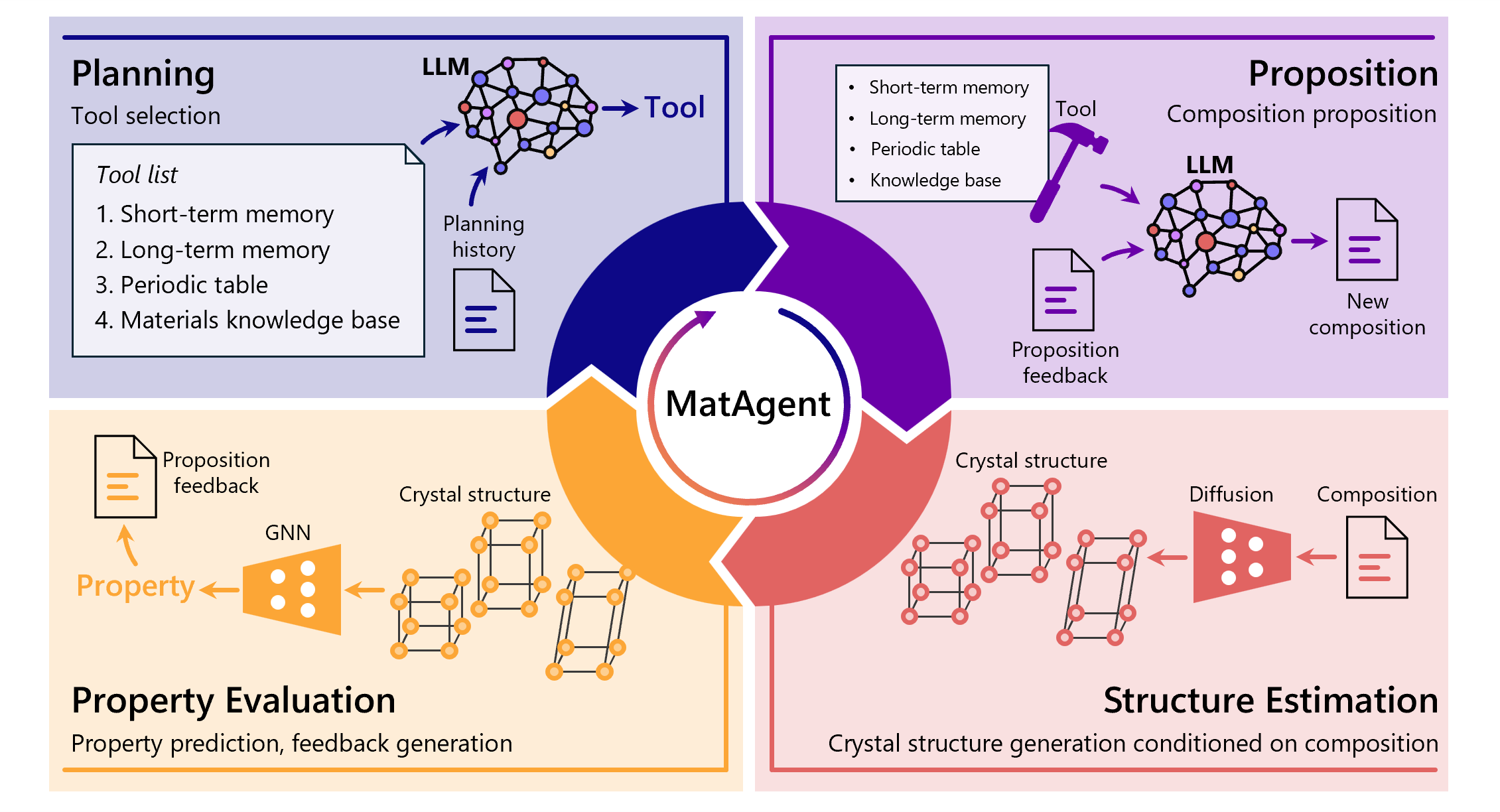}
    \caption{\textbf{The overview of the proposed materials design framework.} The large language model (LLM) generates materials compositions through two sequential stages: Planning and Proposition. During the Planning stage, the LLM selects the appropriate tools (short-term memory, long-term memory, periodic table, and materials knowledge base), providing the natural-language justifications. In the Proposition stage, the LLM proposes revised compositions by reasoning based on insights obtained from these tools. The proposed compositions are then processed by the Structure Estimator (a diffusion model), which generates corresponding three-dimensional crystal structures. These structures are evaluated by the Property Evaluator, and feedback from the Property Evaluator is iteratively provided to the LLM for further refinement.}
    \label{fig:framework}
\end{figure}

An overview of the proposed \framework framework is shown in Figure \ref{fig:framework}. \framework employs an LLM as the central generative engine, leveraging its reasoning capabilities to ensure interpretability in the materials design process. By integrating external tools such as short-term memory, long-term memory, a periodic table, and a materials knowledge base, along with predictive and generative models, \framework facilitates the iterative exploration of a broader materials space. Each iteration of \framework comprises four distinct steps: (1) LLM-driven Planning stage, (2) LLM-driven Proposition stage, (3) generation of three-dimensional crystal structures by the Structure Estimator, and (4) Evaluation of materials properties and the preparation of feedback by the Property Evaluator.

\subsubsection*{LLM as agent}
\begin{figure}[htbp]
    \centering
    \includegraphics[width=\linewidth]{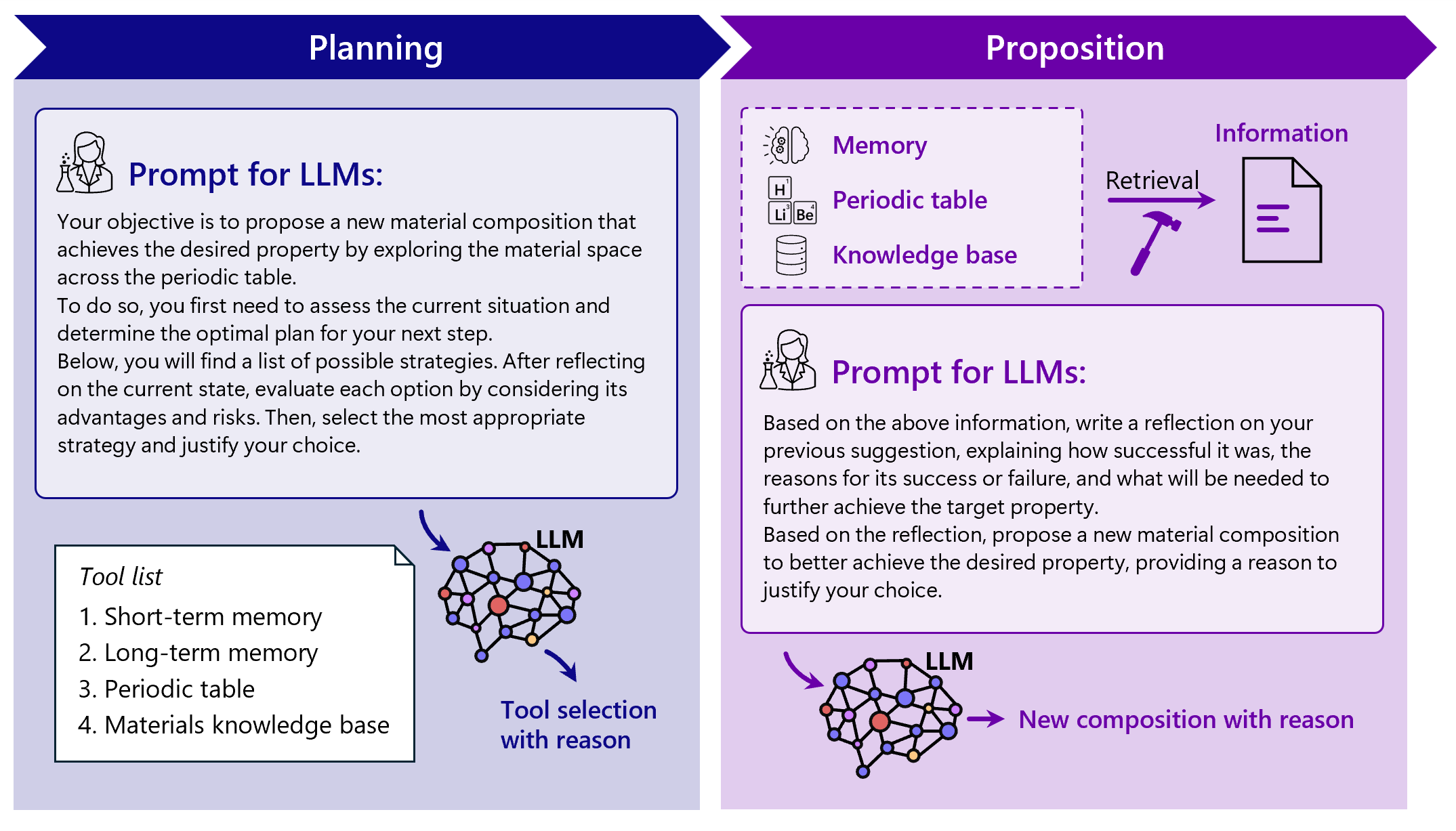}
    \caption{\textbf{Workflow of composition proposal by the LLM.} In the Planning stage, the LLM first analyzes the current context and selects an appropriate tool with explicit reasoning. During the subsequent Proposition stage, information retrieval is performed based on the selected tool. Leveraging this retrieved information, alongside the previously proposed composition and corresponding feedback, the LLM then proposes a new composition, again accompanied by explicit reasoning that ensures interpretability in the materials design process.}
    \label{fig:prompt}
\end{figure}

Recent advancements in LLMs have driven remarkable improvements in their reasoning capabilities, enabling their use in complex problem-solving tasks\cite{boiko2023autonomous,zhang2024honeycomb,bran2024augmenting}. In this study, we leverage such enhanced reasoning capabilities of LLMs for AI-driven materials design. Specifically, the LLM generates new material compositions through a two-stage reasoning process: Planning and Proposition, as shown in Figure \ref{fig:prompt}.

In the Planning stage, the LLM first analyzes the current situation and strategically determines how to proceed with proposing the next material composition. It evaluates the current context, which includes information about recent tool selections and the resulting changes in the properties of previously proposed compositions. Based on this analysis and the given target property value, the LLM selects one of the available tools--—short-term memory, long-term memory, periodic table, or materials knowledge base--—and provides an explicit justification for its choice to guide subsequent composition proposals. The complete prompt template for this Planning stage is provided in the supplementary material \ref{sup:planning}.

In the Proposition stage, relevant information is retrieved based on the tool selected during the Planning stage. Combining this retrieved information with the composition proposed in the previous iteration and the corresponding feedback, the LLM then generates a new composition proposal accompanied by explicit reasoning, providing interpretable insights into the underlying design choices. The complete prompt templates for this Proposition stage corresponding to each tool are shown in the supplementary material \ref{sup:proposition}.

\subsubsection*{Tools}
While LLMs possess extensive general knowledge and fundamental reasoning capabilities, it remains uncertain whether LLMs alone can effectively navigate the complexities inherent in materials design. To enhance the reasoning capabilities of LLMs specifically within materials design and enable effective AI-driven materials design, we aimed to mimic human expert reasoning by integrating strategic tool use into our framework. Specifically, we integrated four distinct tools---short-term memory, long-term memory, the periodic table, and the materials knowledge base---reflecting the human approach of leveraging short-term experience, long-term insights, fundamental knowledge such as the periodic table, and previously accumulated knowledge to facilitate iterative exploration and informed decision-making in materials-design.

The short-term memory recalls the compositions proposed in recent iterations and the corresponding feedback received. The intent of this is to help the LLM understand recent performance trends and revise potentially redundant composition proposals. The long-term memory retrieves not only previously successful compositions but also the associated reasoning processes used by the LLM, thus providing insights into why certain compositions led to favorable outcomes. The periodic table provides elements related to those used in the previous composition, specifically elements within the same group, based on the expectation that such substitutions can fine-tune material properties while preserving compositional plausibility. Lastly, the materials knowledge base is a compiled database that records how material properties change when transitioning from one composition (source composition) to another (target composition). The previously proposed composition is used as a query to search this database, specifically by evaluating its similarity to the entries of source composition. Subsequently, the LLM screens the retrieved entries, selectively filtering for information that appears most relevant and useful for further analysis and decision-making. By strategically utilizing these tools, the framework moves beyond the intrinsic knowledge of the LLM, enhancing its reasoning capabilities and broadening exploration within the materials design space.

\subsubsection*{Structure Estimator}
Although the LLM proposes candidate materials as discrete compositions, accurately evaluating the corresponding material properties typically requires knowledge of three-dimensional crystal structures, including unit cell geometries and atomic coordinates. To address this requirement, the Structure Estimator plays a crucial role in the proposed framework by estimating three-dimensional crystal structures corresponding to the compositions proposed by the LLM.

In this study, we employed a diffusion model\cite{ho2020ddpm, song2021score, jiao2023diffcsp} as the backbone of the Structure Estimator. Diffusion models have successfully been applied to generative tasks within image and language generation fields, and their effectiveness has also been demonstrated in crystal structure generation\cite{Zeni2025, jiao2023diffcsp, yang2024scalable}. We constructed the MP-60 dataset by collecting stable crystal structure data with unit cells containing up to 60 atoms from the Materials Project\cite{jain2013mp} database. Using this dataset, we trained a conditional crystal structure generation model, which generates multiple candidate structures for each given reduced composition with varying numbers of formula units per unit cell ($Z$). This approach is based on the idea that simultaneously generating and evaluating structures with different formula units for the same composition helps identify the most stable unit cell configuration. 

\subsubsection*{Property Evaluator}
The Property Evaluator plays a key role by quantitatively assessing and providing feedback on the material compositions proposed by the LLM. Specifically, it evaluates the physical properties of three-dimensional crystal structures generated by the Structure Estimator. In this study, we constructed a property predictor based on graph neural networks (GNNs), trained on the MP-60 dataset, specifically for predicting the formation energy per atom. The formation energy per atom of all candidate structures is evaluated, and the structure with the lowest formation energy per atom is considered the most stable and selected. Feedback derived from the formation energy predictions is subsequently returned to the LLM, enabling further refinement and improvement of the proposed compositions. A template for generating this feedback is provided in the supplementary material \ref{sup:feedback}.

%% file: body/1_results_2_evaluation.tex
\subsection*{Iterative materials generation guided by target property}

Our proposed framework is designed to iteratively refine material composition proposals through successive feedback cycles, progressively guiding them toward specific target properties. As a representative demonstration, we selected the \textit{formation energy per atom} as the target property, aiming to systematically explore the materials space toward compositions exhibiting desired formation energies. Formation energy serves as an indicator of the chemical bonding strength and stability of materials, and it has frequently been employed to evaluate the performance of generative methods in inverse materials design tasks\cite{xie2022crystal, luo2023symat}. Specifically, we defined target formation energy values based on quantiles derived from the MP-60 dataset. Quantiles were calculated by sorting the formation energy values in ascending order: $-$3.8 eV/atom (1.0\% quantile), $-$3.5 eV/atom (2.5\% quantile), $-$3.0 eV/atom (10\% quantile), $-$2.5 eV/atom (20\% quantile), and $-$1.6 eV/atom (40\% quantile). The distribution of formation energy per atom for the MP-60 dataset is provided in the supplementary material \ref{sup:data-dist}. To evaluate the performance of our proposed framework, we employed OpenAI's GPT-4o (gpt-4o-2024-08-06) and o3-mini (o3-mini-2025-01-31) as the LLM backbones. We conducted 20 independent runs consisting of 15 iterations each, starting from randomly sampled compositions drawn from an existing dataset. GPT-4o is a general-purpose LLM developed by OpenAI, applicable to various types of tasks, while o3-mini is a specialized model optimized for focused reasoning applications with strength in scientific contexts.

\begin{figure}[htbp]
    \centering
    \includegraphics[width=\linewidth]{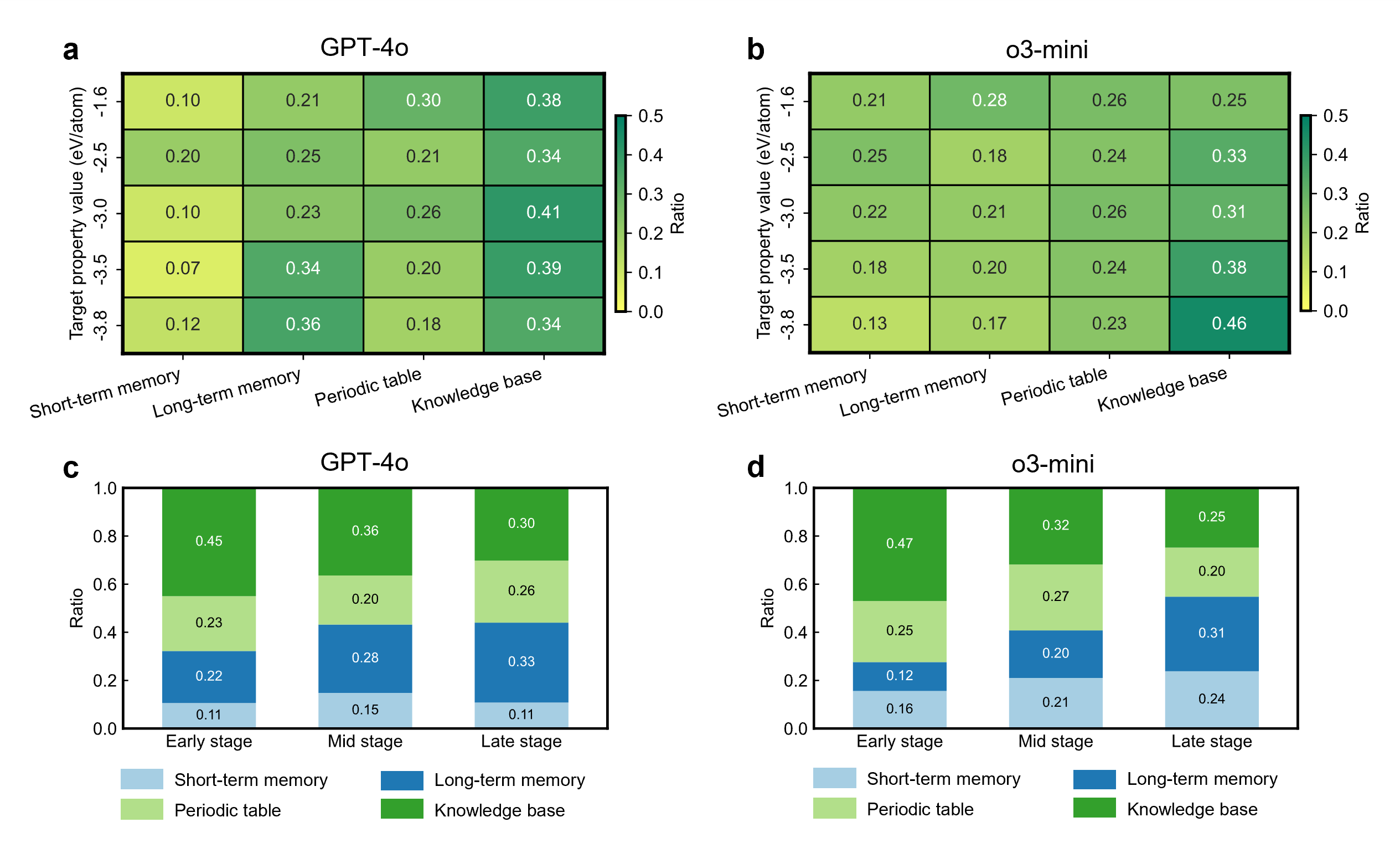}
    \caption{\textbf{Comparison of tool usage patterns during the Planning stage by the LLM.} Heatmaps \textbf{a} and \textbf{b} illustrate the frequency with which each tool (Short-term memory, Long-term memory, Periodic table, and Knowledge base) was selected by the LLM during the Planning stage to guide subsequent material composition proposals. Panel \textbf{a} presents results obtained using GPT-4o, while panel \textbf{b} shows results using o3-mini. Panel \textbf{c} and \textbf{d} further illustrate how tool selection patterns evolved over iterations, categorized into Early (iterations 1--5), Mid (iteration 6--10), and Late stages (iterations 11--15). Panel \textbf{c} corresponds to results with GPT-4o, and panel \textbf{d} corresponds to the results with o3-mini.}
    \label{fig:choice}
\end{figure}

Since the proposed framework employs an iterative approach, in which the LLM strategically selects tools to guide composition proposals, we first analyzed how each LLM selected these tools during the Planning stage. Figure \ref{fig:choice} summarizes the frequency of tool selection for each LLM. Panels \textbf{a} and \textbf{b} in Figure \ref{fig:choice} show the proportions of each tool selected by GPT-4o and o3-mini, respectively, across different target formation energies. Panels \textbf{c} and \textbf{d} further illustrate how the tool-selection patterns evolved over iterations for GPT-4o and o3-mini. In panels \textbf{c} and \textbf{d}, we defined the Early stage as iterations 1--5, the Mid stage as iterations 6--10, and the Late stage as iterations 11--15 out of the total 15 iterations.


GPT-4o and o3-mini showed distinct patterns of tool selection depending on the target formation energy, as illustrated in Figure \ref{fig:choice} \textbf{a} and \textbf{b}. Specifically, GPT-4o frequently selected the materials knowledge base and long-term memory when targeting lower formation energy values (e.g., $-$3.8 eV/atom), whereas at higher values (e.g., $-$1.6 eV/atom), it preferred using the periodic table in addition to the knowledge base. Additionally, GPT-4o showed a relatively low preference for short-term memory. In contrast, o3-mini heavily relied on the materials knowledge base when targeting the challenging formation energy of $-$3.8 eV/atom, but exhibited a more uniform distribution in tool selection when targeting the less challenging value of $-$1.6 eV/atom.

Examining the evolution of tool-selection patterns over iterations, we found that GPT-4o initially showed a strong preference for the knowledge base during the early stage, but gradually reduced its reliance on this tool in later stages. Instead, it increasingly utilized long-term memory to leverage previously successful experiences. A similar trend was observed for o3-mini, which also shifted toward increased use of long-term memory in later stages, accompanied by greater reliance on short-term memory. This suggests that, similar to GPT-4o, o3-mini adapts its tool selection based on the evolving context of the task. This adaptive behavior, utilizing insights from past experiences to inform subsequent selections, can also be observed and interpreted in the reasoning texts generated by the LLM. Examples of these explanations are provided in the supplementary material \ref{sup:reason}.

\begin{figure}[htbp]
    \centering
    \includegraphics[width=0.9\linewidth]{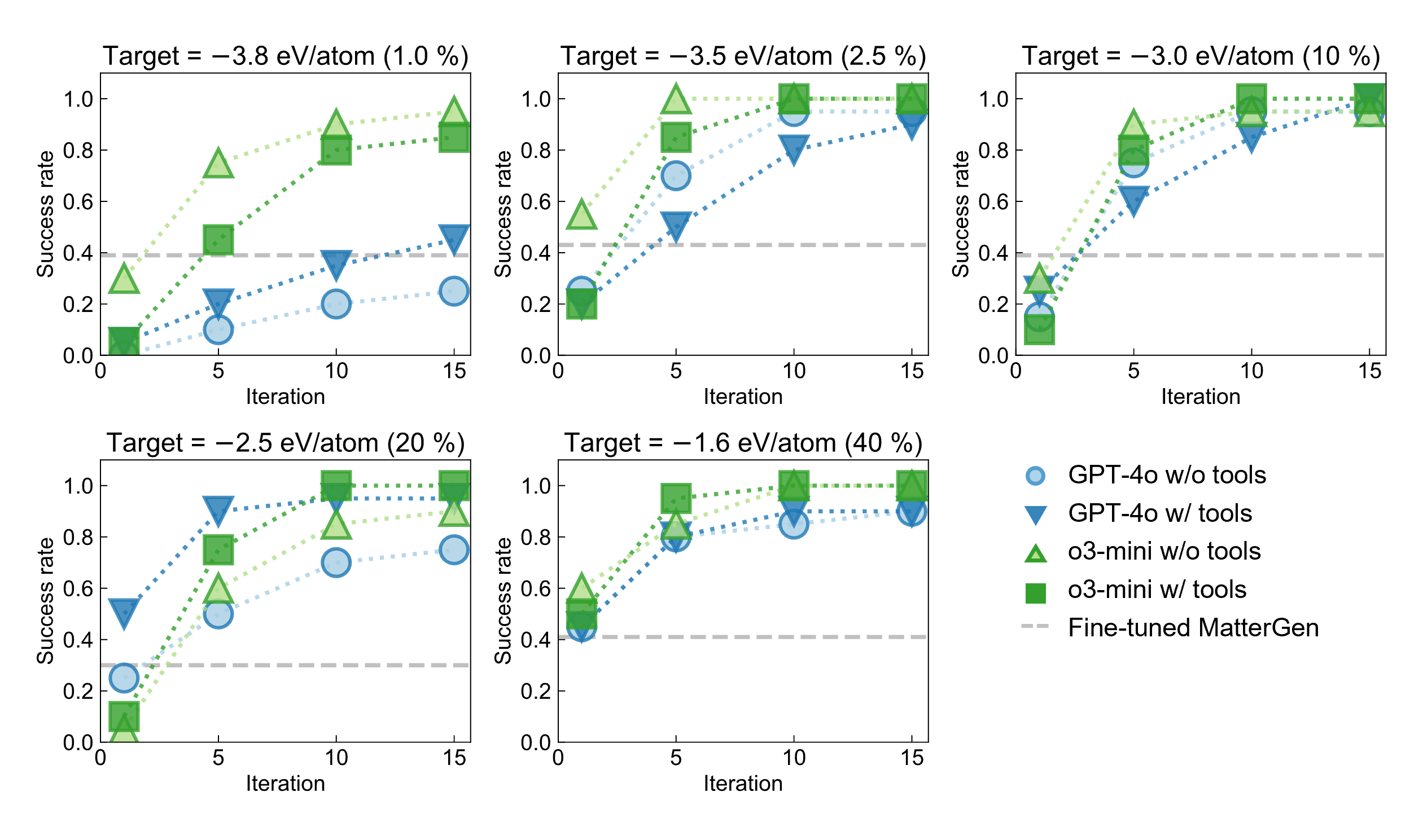}
    \caption{\textbf{Success rate progression over iterations.} Results are shown for the \framework with GPT-4o and o3-mini backbones with and without the use of tool-assisted Planning and Proposition, denoted as GPT-4o w/ tools, GPT-4o w/o tools, o3-mini w/ tools, and o3-mini w/o tools, respectively. Reference success rates obtained using MatterGen\cite{Zeni2025} fine-tuned on formation energy per atom in MP-20 dataset are also included.}
    \label{fig:SR}
\end{figure}

As previously discussed, the pattern of tool selection during the Planning stage varied depending on the LLM backbone used. To further explore how each LLM differs in its ability to propose materials that meet specific target properties, we analyzed the Proposition stage, in which the LLM retrieves information using the tools selected during the Planning stage, derives insights, and generates new composition proposals along with natural-language explanations. Success rates for each scenario were calculated based on these proposals, as shown in Figure \ref{fig:SR}. Here, a ``successful'' run at a given iteration is defined as a case in which at least one composition proposed up to that iteration achieved a predicted property value within ±0.25 eV/atom of the target. We adopted this relatively broad threshold to account for inherent prediction errors arising from the Structure Estimator and Property Evaluator. While our proposed framework was designed to utilize external tools during the Planning and Proposition stages, we also conducted comparison experiments without tool-assisted Planning and Proposition (denoted as GPT-4o w/o tools and o3-mini w/o tools). The prompt template used for these experiments is provided in the supplementary material \ref{sup:baseline}. Figure \ref{fig:SR} also includes the success rates obtained using MatterGen, a diffusion-based generative model that simultaneously generates lattice vectors, atomic coordinates, and atomic species. As proposed in the original MatterGen study\cite{Zeni2025}, classifier-free (CF) guidance\cite{ho2022classifier} was employed to enable inverse design toward specific target formation energies, with a diffusion guidance scale of 2.0. We fine-tuned MatterGen on the MP-20 dataset, a curated dataset previously utilized for crystal structure generation tasks\cite{xie2022crystal, luo2023symat, jiao2023diffcsp, Zeni2025}, to perform CF-guided diffusion conditioned on target formation energies. For MatterGen, we generated 256 candidate structures and calculated the success rate based on the percentage of structures whose predicted formation energy fell within $\pm$0.25 eV/atom of the target value. Since both MatAgent and MatterGen rely on the same property predictor to estimate formation energies, the resulting success rates are influenced by the predictor's accuracy and potential biases, and thus may include some degree of error. Additionally, the success rates for both MatAgent and MatterGen were calculated without any post-processing steps such as structural relaxation or optimization. It should be noted that while MatterGen generates materials via a single diffusion process, \framework utilizes a similar diffusion model in the Structure Estimator, resulting in increasing computational costs as iteration progresses. For the initialization of \framework in this experiment, initial compositions were randomly sampled from the MP-20 dataset.

In our proposed framework, it is expected that material properties will progressively approach the specified target values as iteration proceeds. Consistent with this expectation, Figure \ref{fig:SR} shows an increase in the success rate as the number of iterations increased, confirming the effectiveness of iterative refinement toward the targeted formation energies. It was also observed that MatterGen typically achieved higher success rates at early iterations compared to both GPT-4o and o3-mini, indicating its effectiveness in rapidly generating materials close to target formation energies. The results indicate that GPT-4o w/ tools achieved higher success rates compared to GPT-4o w/o tools especially when targeting formation energies of $-$3.8 eV/atom and $-$2.5 eV/atom. However, for the most challenging target property corresponding to the lowest 1.0 \% quantile ($-$3.8 eV/atom), both GPT-4o w/ tools and w/o tools exhibited relatively low success rates, reflecting the difficulty of identifying compositions in lowest-formation-energy regions. On the other hand, o3-mini consistently achieved high success rates even at challenging target values (e.g., $-$3.8 eV/atom and $-$3.5 eV/atom). Additionally, we observed that o3-mini w/o tools demonstrated higher success rates compared to o3-mini w/ tools, the reason for which will be discussed later in this section.

\begin{figure}[htbp]
    \centering
    \includegraphics[width=0.9\linewidth]{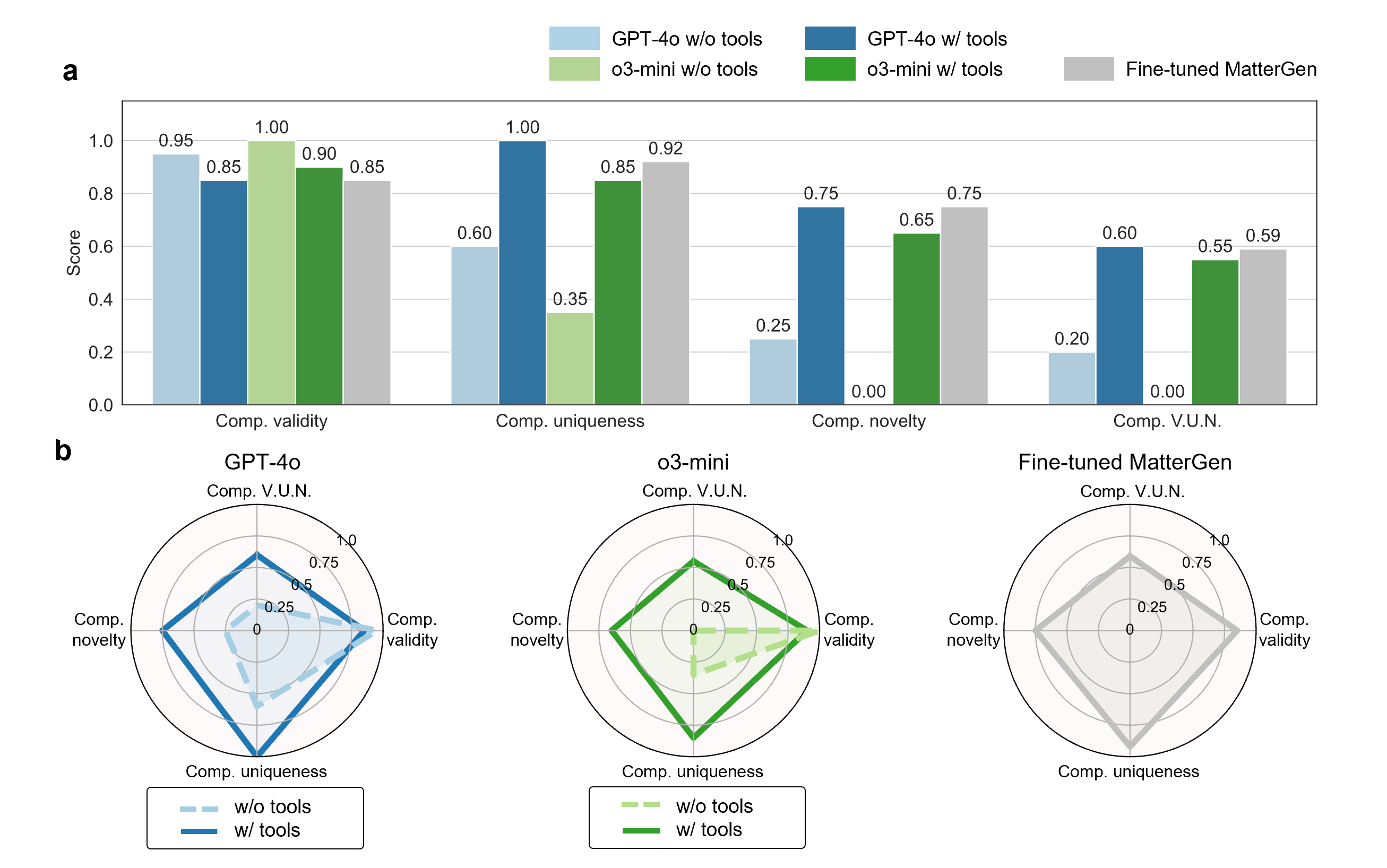}
    \caption{\textbf{Calculated compositional metrics.} \textbf{a} Comparison of compositional validity, uniqueness, novelty, and V.U.N. scores across five different approaches. \textbf{b} Radar charts summarizing these compositional metrics for each LLM backbone.}
    \label{fig:comp}
\end{figure}

Although the high success rates achieved by GPT-4o and o3-mini indicate their capability to generate material compositions close to target properties, these success rates alone may not fully reflect the practical usefulness or broader effectiveness of the framework in materials exploration. In practice, compositions that are chemically or physically invalid or already well-known typically have limited usefulness. Therefore, the compositional validity, uniqueness, novelty, and overall V.U.N. score of the generated materials compositions when targeting $-$3.8 eV/atom were evaluated as shown in Figure \ref{fig:comp}. Here, validity refers to the proportion of compositions assessed valid according to SMACT\cite{Davies2019smact}; uniqueness denotes the proportion of unique compositions among candidate compositions; novelty indicates the proportion of compositions not present in the MP-20 training set; and the V.U.N. score represents the proportion of compositions that are simultaneously valid, unique, and novel. Here, it is important to note that a direct comparison of novelty scores between \framework and MatterGen may not be entirely fair. While MatterGen was fine-tuned on the MP-20 dataset, the LLMs in \framework were not exclusively trained on this dataset, and their broader pretraining may influence the novelty evaluation. Figure \ref{fig:comp} \textbf{a} shows the comparison of each metric across different approaches, and Figure \ref{fig:comp} \textbf{b} presents radar charts summarizing these compositional metrics for each LLM backbone. For reference, the SMACT validity of compositions in the MP-20 dataset is 0.89, suggesting that achieving a validity score close to 1.0 does not necessarily imply improved quality or practical usefulness. All metrics were evaluated based on the structures identified by the Property Evaluator as closest to the target property value within each independent run. In the evaluation of MatterGen, all compositions generated through CF-guided generation were included in the assessment.

When GPT-4o was employed as the backbone LLM, clear differences emerged between the w/o tools and w/ tools scenarios. Although compositional validity remained high in the w/o tools scenarios, uniqueness and novelty scores were relatively low. Upon inspecting the specific compositions proposed in this case, we frequently encountered well-known compounds such as \ce{TiO2} and \ce{SrTiO3}. In contrast, incorporating tool-assisted Planning and Proposition significantly enhanced uniqueness and novelty, resulting in an increased V.U.N. score from 0.20 (GPT-4o w/o tools) to 0.60 (GPT-4o w/ tools).

A similar, yet even clearer trend was observed when using o3-mini as the backbone LLM, as confirmed by the shapes of the radar chart in Figure \ref{fig:comp} \textbf{b}. Specifically, o3-mini w/o tools exhibited a compositional validity of 1.0 but achieved a novelty score of 0.0, resulting in a compositional V.U.N. score of 0.0. In contrast, o3-mini w/ tools significantly increased uniqueness and novelty, yielding a V.U.N. score of 0.55, which is comparable to the performance obtained with MatterGen. In the case of o3-mini w/o tools, about 60 \% of candidate compositions were \ce{HfO2}, suggesting that the LLM heavily relied on the inherent knowledge within o3-mini, specifically the prior information that \ce{HfO2} is a stable compound. Although o3-mini w/o tools exhibited a high success rate when targeting formation energy of $-$3.8 eV/atom as shown in Figure \ref{fig:SR}, this was likely due to its heavy reliance on proposing well-known, stable compounds such as \ce{HfO2}, rather than effectively exploring new compositions.

\begin{figure}[htbp]
    \centering
    \includegraphics[width=\linewidth]{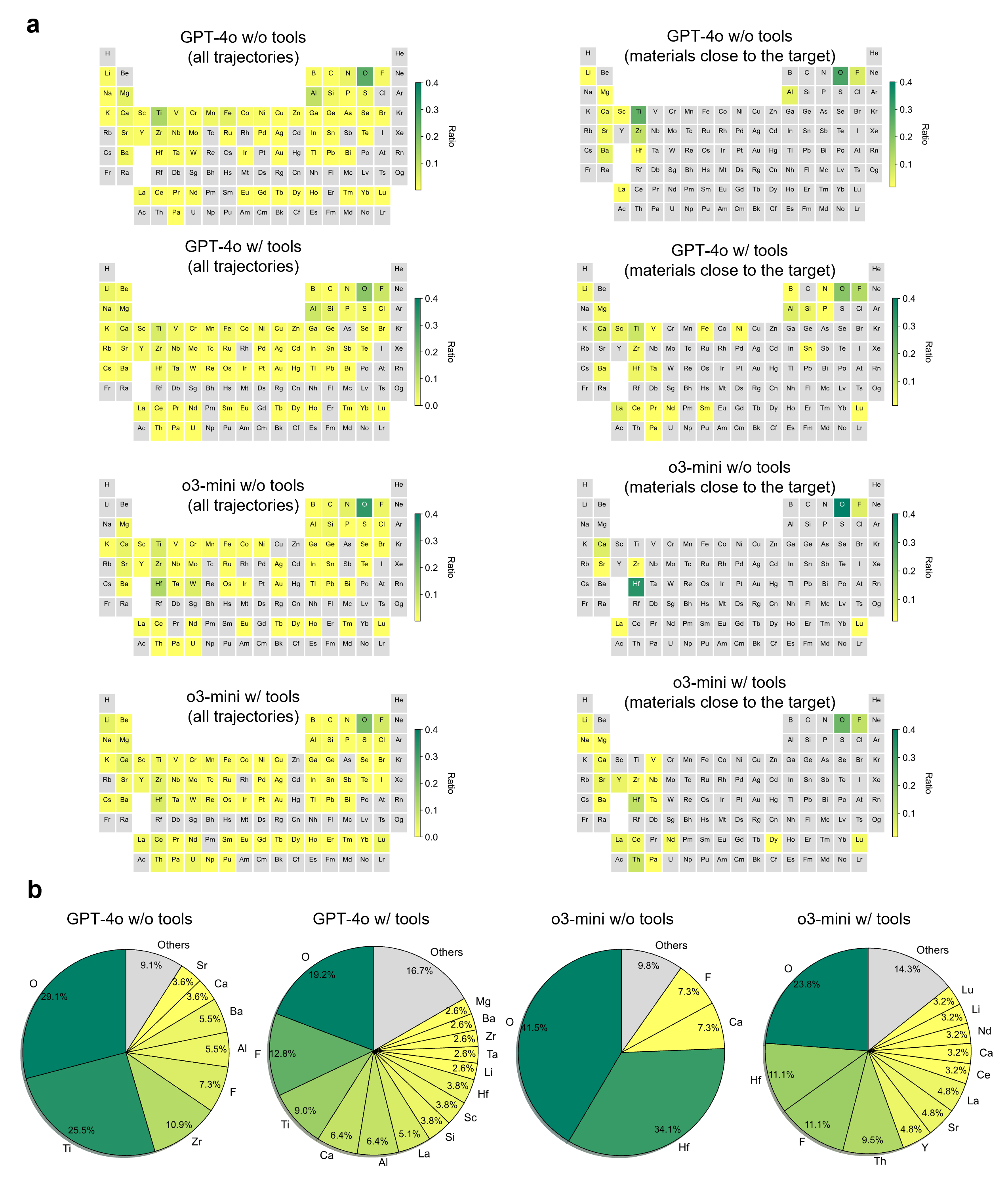}
    \caption{\textbf{Elemental distributions of proposed compositions.} \textbf{a} Elemental distributions calculated from compositions proposed across 15 iterations in 20 independent runs targeting a formation energy of -3.8 eV/atom. The left panels show frequencies from all proposed compositions; the right panels show frequencies from compositions closest to the target. Rows one to four correspond to GPT-4o w/o tools, GPT-4o w/ tools, o3-mini w/o tools, and o3-mini w/ tools, respectively. \textbf{b} Pie charts showing elemental proportions of compositions closest to the target, sorted by frequency.}
    \label{fig:elem_dist}
\end{figure}

To quantitatively understand which elements were frequently proposed, we analyzed the elemental distributions in the compositions generated across all runs. The results are presented in Figure \ref{fig:elem_dist}. In Figure \ref{fig:elem_dist} \textbf{a}, the left-column panels show elemental frequencies calculated from all proposed compositions, while the right-column panels show frequencies calculated only from compositions identified as closest to the target property. Rows one through four correspond to GPT-4o w/o tools, GPT-4o w/ tools, o3-mini w/o tools, and o3-mini w/ tools, respectively. Figure \ref{fig:elem_dist} \textbf{b} shows pie charts illustrating elemental proportions of compositions closest to the target, sorted by descending frequency.

Comparing GPT-4o w/o tools and GPT-4o w/ tools, we observed that incorporating tool-assisted Planning and Proposition significantly expanded the diversity of the explored compositional space. Specifically, among the compositions closest to the target property, GPT-4o w/o tools primarily proposed materials dominated by \ce{Ti} and \ce{O}, while GPT-4o w/ tools led to a more diverse range of compositions, frequently involving elements such as \ce{F}, \ce{Ca}, and \ce{Al}. A similar expansion in the variety of explored elements was observed for o3-mini when tool-assisted Planning and Proposition were employed. Notably, the elements frequently appearing in compositions closest to the target property differed between GPT-4o and o3-mini. In particular, o3-mini w/o tools primarily suggested compositions containing a limited set of elements, such as \ce{Hf} and \ce{O}, resulting in high success rates but low compositional novelty. In contrast, o3-mini w/ tools introduced a broader range of elements, including \ce{F}, \ce{Th}, and \ce{Y}, enhancing both compositional diversity and novelty.

Combining the success rates shown in Figure \ref{fig:SR} and the compositional metrics presented in Figure \ref{fig:comp} and Figure \ref{fig:elem_dist}, clear differences emerged depending on the use of tool-assisted Planning and Proposition. Without tool assistance, the LLMs frequently proposed well-known, typical compositions, achieving high success rates but exhibiting limited compositional novelty. In contrast, incorporating tool-assisted Planning and Proposition allowed the LLMs to significantly improve compositional uniqueness and novelty while maintaining comparably high success rates, highlighting the advantages of integrating external tools into the LLM-driven reasoning process.

To investigate the individual contributions of each tool to the observed improvements, we conducted additional experiments using GPT-4o as the LLM backbone, evaluating each tool independently. The resulting compositional metrics, number of elements within candidate materials, and success rates are summarized in Table \ref{tab:tool}.

\begin{table}[htbp]
    \centering
    \begin{tabular}{|l|c|c|c|c|c|}
    \hline
    Method & Comp. validity & Comp. uniqueness & Comp. novelty & Num. of elements & Success rate \\
    \hline
    w/o tools & 0.95 & 0.60 & 0.25 & 13 & 0.25 \\
    \hdashline
    w/ Short-term memory & 0.95 & 0.85 & 0.60 & 19 & 0.30\\
    w/ Long-term memory & 0.90 & 1.00 & 0.85 & 25 & 0.25\\
    w/ Periodic table & 0.80 & 1.00 & 0.60 & 36 & 0.10\\
    w/ Knowledge base & 0.95 & 1.00 & 0.70 & 24 & 0.65\\
    \hdashline
    w/ tools & 0.85 & 1.00 & 0.75 & 27 & 0.45\\
    \hline
    \end{tabular}
    \caption{Evaluation metrics for materials generated using each tool independently.}
    \label{tab:tool}
\end{table}

While Figure \ref{fig:choice} shows that the materials knowledge base was frequently selected by the LLM, Table \ref{tab:tool} suggests that the use of this knowledge base may contribute to proposing compositions that are both unique and associated with high success rates. Additionally, long-term memory appears beneficial in enhancing compositional uniqueness and novelty, while the periodic table likely aids in exploring novel elemental spaces.

%% file: body/1_results_3_init.tex
\subsection*{Natural language integration to initialization}

Up to this point, initial compositions within the framework have been randomly selected from the MP-20 dataset, demonstrating that this approach allows compositions to effectively navigate toward a specified target property. However, a more intuitive and user-friendly alternative is to allow human users to directly initiate the composition generation process through natural-language prompts. To achieve this, two distinct methods were implemented: one where the LLM proposes an initial composition based on a provided prompt, and another that employs a retriever trained to identify relevant compositions from a database using natural-language queries.

\begin{figure}[htbp]
    \centering
    \includegraphics[width=0.85\linewidth]{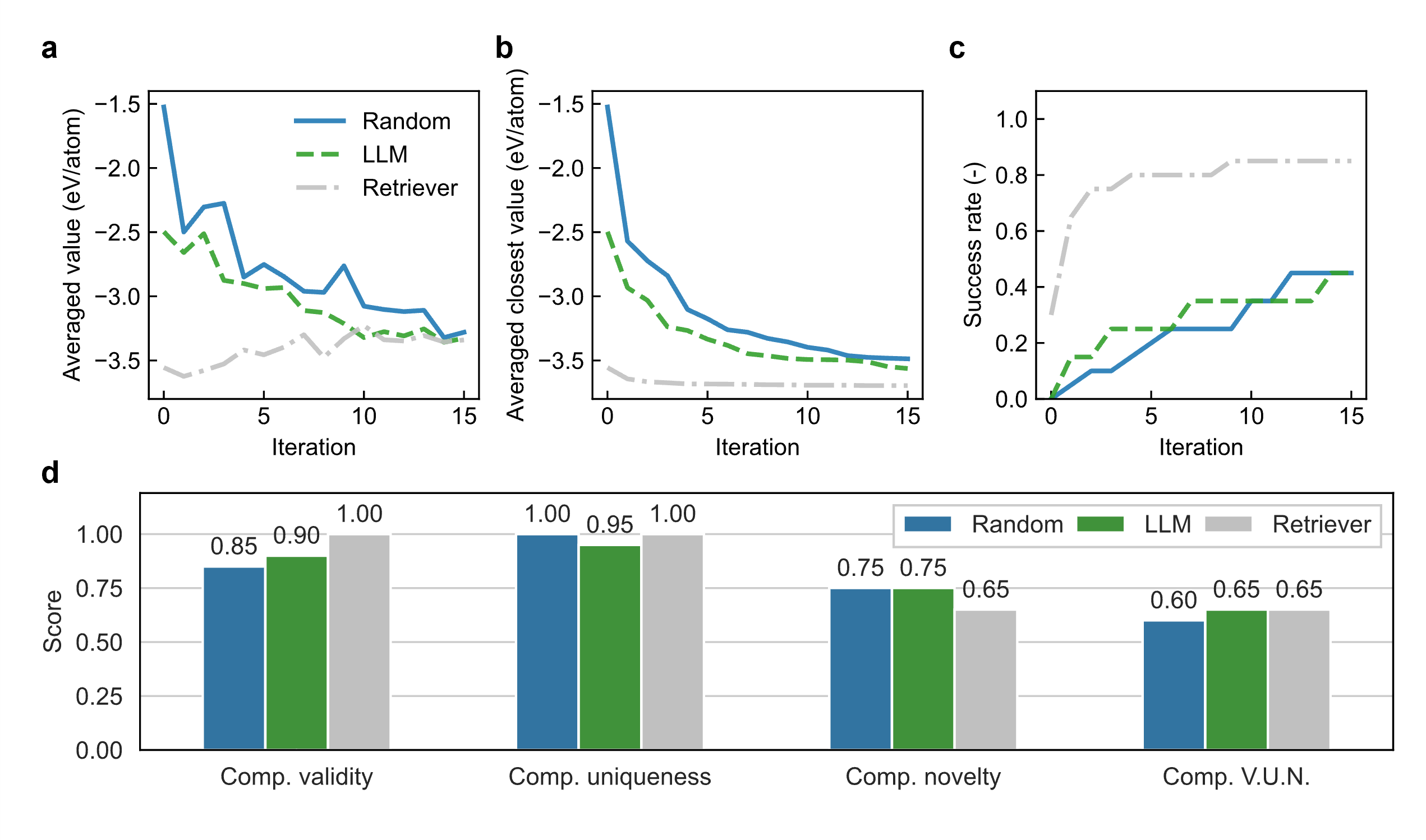}
    \caption{\textbf{Comparison of three initialization methods (Random, LLM, Retriever).} Target formation energy was set to -3.8 eV/atom. \textbf{a} Average property values at each iteration across 20 independent runs. \textbf{b} Average of the property values closest to the target achieved up to each iteration, calculated across 20 independent runs. \textbf{c} Success rates calculated at each iteration, averaged over 20 independent runs. \textbf{d} Compositional metrics (validity, uniqueness, novelty, and V.U.N. score) for compositions identified as closest to the target.}
    \label{fig:init}
\end{figure}

Figure \ref{fig:init} compares three different initialization methods used to reach a target formation energy of $-$3.8 eV/atom. Panel \textbf{a} shows how the average property values evolve across iterations, averaged over 20 independent runs. Panel \textbf{b} presents the average of the closest property values to the target observed in all preceding iterations. Panel \textbf{c} presents success rates over iterations, where the success rate is defined as achieving a property value within $\pm$0.25 eV/atom of the target. Finally, panel \textbf{d} summarizes compositional metrics (validity, uniqueness, novelty, and the V.U.N. score) for the compositions successfully identified as being close to the target value.

From Figure \ref{fig:init} \textbf{a}, it is clear that initialization methods utilizing the LLM and Retriever yield initial material compositions closer to the target property than the Random initialization method. In particular, the Retriever-based initialization, which provides material compositions with property values closer to the target, facilitates a more efficient and targeted exploration around the desired property region. As a result, as shown in Figure \ref{fig:init} \textbf{c}, this approach achieves a higher overall success rate. Although the compositions generated using the Retriever exhibit slightly lower novelty, the compositional V.U.N. scores remain comparable across all initialization methods. These results underscore the practical advantages of Retriever-based initialization in enabling intuitive and effective material generation toward target properties.

%% file: body/1_results_4_constraint.tex
\subsection*{Constrained generation with natural language}

In practical materials design, it is common to impose specific constraints reflecting environmental considerations, resource availability, or recycling requirements. For instance, one may require materials to consist solely of environmentally friendly elements or to be easily recyclable using commonly available recycling processes. However, conventional generative approaches in materials design have struggled to flexibly integrate such qualitative constraints, typically being restricted to numeric or categorical conditions. Since the proposed framework employs an LLM as its central generative engine, it is straightforward to incorporate diverse constraints expressed in natural language into the framework. To evaluate this capability, we examined the performance of our proposed framework under several practical constraints:
\begin{enumerate}
    \item Exclusion of environmentally damaging elements: Materials generation was performed while explicitly excluding \ce{Pb}, \ce{Hg}, \ce{Cd}, and \ce{Cr}. The constraint was communicated via the prompt: ``Here, do not include \ce{Pb}, \ce{Hg}, \ce{Cd}, \ce{Cr} in any proposed compositions.''
    \item Exclusion of actinide elements: Generation was carried out excluding actinide elements, with the prompt: ``Here, do not include the actinides in any proposed compositions.''
    \item Restriction to non-metal elements: Generation was constrained to compositions containing only non-metal elements, with the prompt: ``Here, only propose compositions consisting of non-metal elements.''
\end{enumerate}

\begin{figure}
    \centering
    \includegraphics[width=0.95\linewidth]{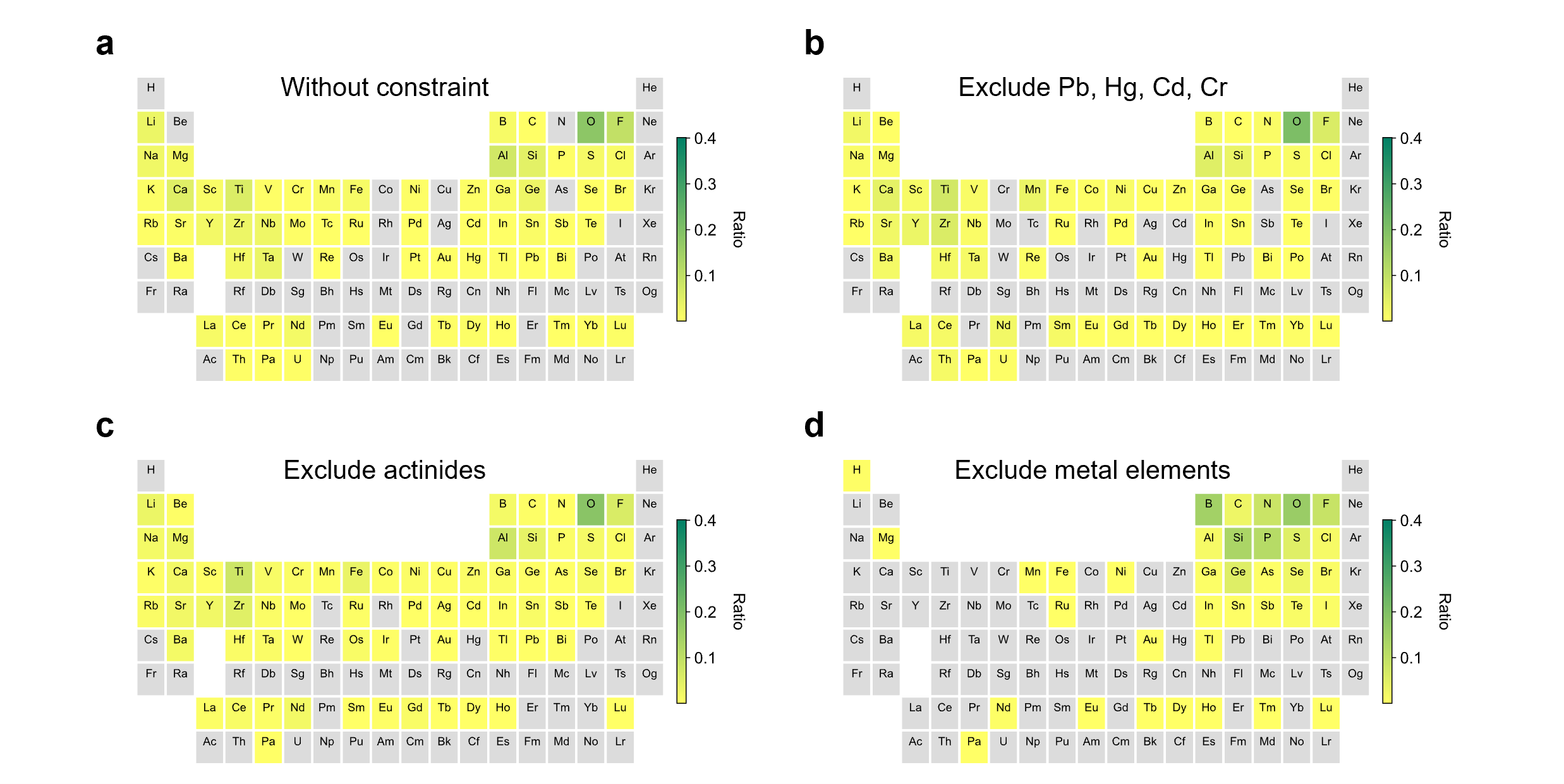}
    \caption{\textbf{Comparison of elemental distributions explored by the framework under different additional prompts.} \textbf{a} Elemental distribution without any additional constraints. \textbf{b} Elemental distribution under the constraint excluding \ce{Pb}, \ce{Hg}, \ce{Cd}, and \ce{Cr}. \textbf{c} Elemental distribution under the constraint excluding actinides elements. \textbf{d} Elemental distribution under the constraint restricting compositions to non-metallic elements only.}
    \label{fig:constraint}
\end{figure}

Introducing constraints is expected to alter the distribution of elements proposed by the LLM. To evaluate this effect, we analyzed elemental distributions of the generated material compositions under different conditions, as shown in Figure \ref{fig:constraint}. Panel \textbf{a} illustrates the elemental distributions without any additional constraints, while panels \textbf{b}--\textbf{d} present the results obtained under specific constraints. Each distribution represents the ratio of elements contained in the proposed compositions generated by the LLM, aggregated over 10 independent runs—each consisting of 15 iterations, using GPT-4o as the LLM backbone. For the generation experiments, the target property values were set to $-$3.8 eV/atom for unconstrained generation and constrained generation under Constraints 1 and 2, while Constraint 3 had a target property value of $-$2.5 eV/atom.

As shown in Figure \ref{fig:constraint} \textbf{a}, in the absence of constraints, compositions containing diverse elements across the periodic table, including metals, metalloids, non-metals, and rare-elements, were generated. In contrast, Figure \ref{fig:constraint} \textbf{b} confirms the successful exclusion of Pb, Hg, Cd, and Cr, with none of these elements appearing in the generated compositions. Similarly, Figure \ref{fig:constraint} \textbf{c} demonstrates effective exclusion of actinide elements, showing nearly complete elimination of these elements from the compositions. Figure \ref{fig:constraint} \textbf{d} highlights that generated compositions were predominantly composed of non-metal elements such as \ce{O}, \ce{B}, \ce{Si}, \ce{P}, \ce{F}, \ce{N}, and \ce{Ge}, reflecting adherence to the imposed constraint.

Quantitatively, the proportion of compositions excluding \ce{Pb}, \ce{Hg}, \ce{Cd}, and \ce{Cr} increased from 0.96 without constraints to 1.00 with Constraint 1. Under Constraint 2, the proportion of compositions excluding actinides improved from 0.97 to 0.99. Furthermore, Constraint 3 significantly raised the proportion of compositions exclusively consisting of non-metal elements from 0.00 to 0.83. These results clearly demonstrate the capability of the proposed LLM-based framework to reliably and effectively incorporate practical, natural-language-defined constraints into generative materials design tasks.

%% file: body/2_discussion.tex
\section*{Discussion}

In this work, we present \frameworkws, an LLM-driven generative framework for inorganic materials design, developed to iteratively propose and refine materials toward specific target properties. \framework leverages explicit, interpretable reasoning capabilities of the LLM to propose promising material compositions. Inspired by the reasoning process of human experts, \framework integrates external tools---short-term memory, long-term memory, the periodic table, and a materials knowledge base---to extend beyond the inherent knowledge of the LLM and facilitate exploration across a broader materials space. Compositions proposed by the LLM are subsequently evaluated by two complementary modules: a Structure Estimator employing a diffusion model, and a Property Evaluator employing a GNN-based prediction model. These modules provide precise feedback, guiding the LLM to iteratively refine and improve the generated compositions, effectively steering the exploration toward materials with user-specified target properties.

The proposed framework demonstrated its capability to effectively guide the generation of inorganic materials toward specified target properties, exemplified in the task of generating materials with desired formation energies. Although compositions generated without external tools exhibited limited compositional diversity and novelty, incorporating external tools significantly enhanced the uniqueness and novelty of proposed compositions while maintaining compositional validity. These results highlight that integrating external resources to mimic human-expert reasoning processes enables the framework to surpass the inherent limitations of the LLM alone, thus facilitating broader exploration and practical applicability in materials discovery tasks. Furthermore, by systematically evaluating initialization methods utilizing natural-language prompts, we demonstrated the effectiveness of a Retriever-based initialization step, highlighting the practical advantages of integrating natural language into materials design workflows. Finally, our results underscore the advantages of employing LLMs as generative engines, as they allow for highly flexible constrained generation and improved efficiency in exploring materials with user-specified conditions.

One limitation of the current study is that the proposed framework focuses exclusively on guiding materials generation toward a single target property--formation energy. In practical materials discovery tasks, it is often necessary to simultaneously optimize multiple materials properties, necessitating more sophisticated property evaluators capable of accurately predicting diverse properties and more complex feedback mechanisms. Therefore, future work should focus on extending the framework to effectively guide the LLM toward practical multi-objective materials design. Additionally, the current framework does not evaluate the correctness of reasoning provided by the LLM. Extending the framework to incorporate a human-in-the-loop approach, wherein human experts provide feedback to ensure accurate reasoning, would be important for developing a more effective framework. Recent studies have also highlighted the potential of symmetry-aware generative model for crystal structure generation\cite{jiao2024space,levy2025symmcd, cao2024space}. Incorporating symmetry-aware models into the Structure Estimator of our framework represents a promising direction for future development, potentially enabling the generation of higher-quality crystal structures with improved structural fidelity. The performance of this framework strongly depends on the effectiveness of the Structure Estimator and Property Evaluator. Consequently, it can directly benefit from, and further advance with, the development of predictive and generative models in the field of materials science.

A promising direction for future research is to integrate synthesizability considerations into the \framework framework. By leveraging the text-based reasoning capabilities of LLMs, extensive knowledge derived from textbooks, scientific literature, and other comprehensive sources can be effectively incorporated. This integration would allow the LLM-based framework to autonomously assess and reason about the practical synthesizability of proposed materials. Developing such a framework that seamlessly combines extensive synthesis-related knowledge with efficient exploration of materials spaces represents a significant advancement toward practical autonomous materials discovery systems.

%% file: body/3_methods.tex
\section*{Methods}
\subsection*{Tools}
\subsubsection*{Short-term memory retrieval}
In short-term memory retrieval, information regarding recent composition proposals, the reasoning behind these proposals, and their corresponding feedback are retrieved. In this study, we retrieve information about the three most recent proposals.

\subsubsection*{Long-term memory retrieval}
In long-term memory retrieval, information is retrieved regarding past proposals that successfully generated compositions close to the target property, including the reasoning behind these proposals. In this study, information about the top three such proposals is retrieved.

\subsubsection*{Periodic table retrieval}
In periodic table retrieval, elements belonging to the same group as each element included in the previously proposed composition are retrieved and listed.

\subsubsection*{Knowledge base retrieval}
In knowledge base retrieval, information is retrieved from a pre-constructed knowledge base. To construct this knowledge base, 10,000 random composition pairs were generated from the MP-20 dataset, and GPT-4o was used to analyze and explain why the material properties change when one composition transitions into another. The prompt template used for generating the explanation is provided in supplementary material \ref{sup:transition}. The knowledge base consists of 10,000 data entries, each containing a pair of compositions, their corresponding properties, and GPT-4o-generated explanations. During retrieval, the previously proposed composition is first used as a query to identify the five most similar source compositions from the knowledge base based on compositional similarity. The compositional similarity is evaluated by vectorizing compositions using the CBFV\cite{CBFV} package. These five retrieved candidate entries are then ranked by the LLM, and the top three entries judged by the LLM to be most relevant and useful for the current task are ultimately selected.

\subsection*{Dataset}
To curate the MP-60 dataset for training Structure Estimator and Property Evaluator in this study, we collected crystal structure data from the Materials Project\cite{jain2013mp} database. Structures were restricted to those containing up to 60 atoms per unit cell, with an energy above hull of 0.02 eV/atom or less. Additionally, we excluded structures consisting exclusively of gaseous elements and those with lattice vectors exceeding 20 \AA. After applying these criteria, we obtained a final dataset comprising 48,323 structures, which was subsequently divided into training, validation, and test sets using a 6:2:2 ratio. 

\subsection*{Structure Estimator}
The Structure Estimator is responsible for generating three-dimensional crystal structures given a specific material composition. As the backbone model for this component, we employed DiffCSP\cite{jiao2023diffcsp}, a diffusion-based crystal structure prediction model, trained on the MP-60 dataset to conditionally generate structures based on the provided compositions. For each given reduced composition, we simultaneously generate five crystal structures for formula units ranging from 1 to 4. Additionally, to ensure the total number of atoms in each crystal structure did not exceed 34, we adjusted the maximum formula units accordingly. 

\subsection*{Property Evaluator}
The Property Evaluator assesses the compositions and crystal structures proposed by the LLM, providing feedback that guides subsequent composition proposals. Specifically, it employs iComFormer\cite{yan2024complete}, a GNN-based transformer model trained on formation energy per atom, as its backbone for predicting material properties. Among candidate structures generated by the Structure Estimator, the structure with the lowest predicted formation energy per atom is considered the most stable and selected accordingly. Furthermore, if the format of a proposed composition is inadequate, feedback is provided to the LLM to prompt corrections and improvements in subsequent iterations. Detailed descriptions of the feedback format are provided in supplementary material \ref{sup:feedback}.

\subsection*{Initialization of composition}
\subsubsection*{Random sampling}
In the random initialization method, initial compositions are randomly sampled from the MP-20 dataset.

\subsubsection*{LLM-based sampling}
In the LLM-based initialization method, the initial composition is proposed directly by the LLM. Specifically, the LLM is prompted using a request in natural language, such as: "I am looking to design a material with a formation energy per atom of \textbf{<target>} eV/atom. Could you suggest one possible material composition?", where \textbf{<target>} denotes the target property value, specifically the desired formation energy value.

\subsubsection*{Retriever-based sampling} 
For the Retriever-based initialization method, we trained a Retriever using contrastive learning\cite{chen2020simple}, employing the T5\cite{raffel2023exploring} as a text encoder, following the approach introduced in MatExpert\cite{ding2024matexpert}. Specifically, we performed contrastive learning by encoding both the property descriptions and their corresponding compositional descriptions (chemical formulas) of materials from the MP-20 dataset using the T5 encoder. This enables the retriever to effectively select relevant initial compositions based on natural-language queries describing desired material properties.

%% file: body/4_appendix_1.tex
\begin{center}
{\Large \textbf{\textsf{Supplementary material}}}
\end{center}

\section{Prompt templates}\label{sup:prompt}

\subsection{Prompt template for the Planning stage}\label{sup:planning}
\begin{figure}[htbp]
    \centering
    \includegraphics[width=0.85\linewidth]{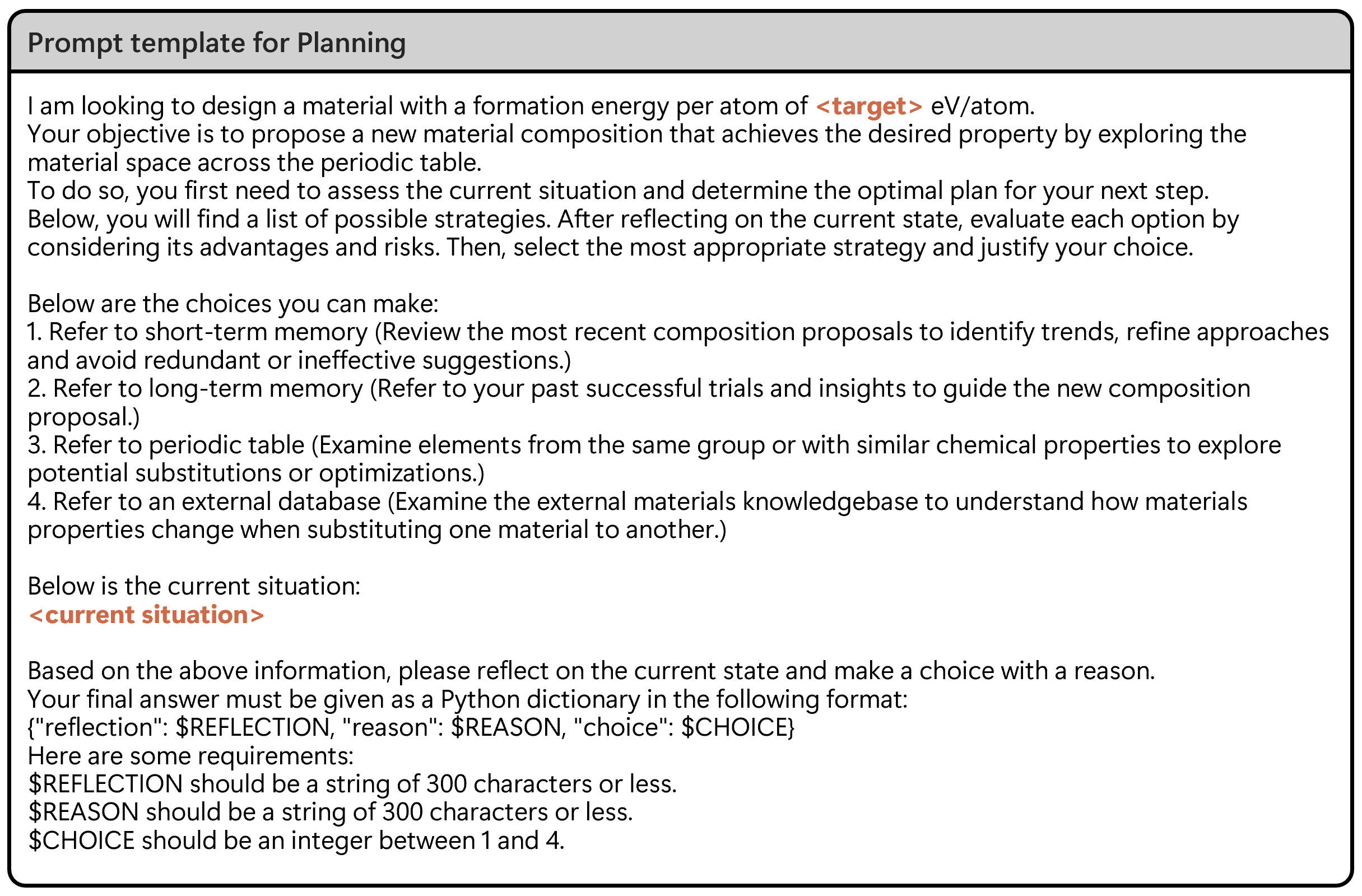}
    \caption{\textbf{Prompt template for Planning stage.} \textbf{<target>} and \textbf{<current situation>} are the variables. Here, \textbf{<current situation>} includes historical information regarding previously selected tools and how the material properties changed as a result of those selections.}
    \label{fig:planning}
\end{figure}

Prompt template used during the Planning stage is shown in Figure \ref{fig:planning}.

\subsection{Prompt templates for the Proposition stage}\label{sup:proposition}
\begin{figure}[htbp]
    \centering
    \includegraphics[width=0.85\linewidth]{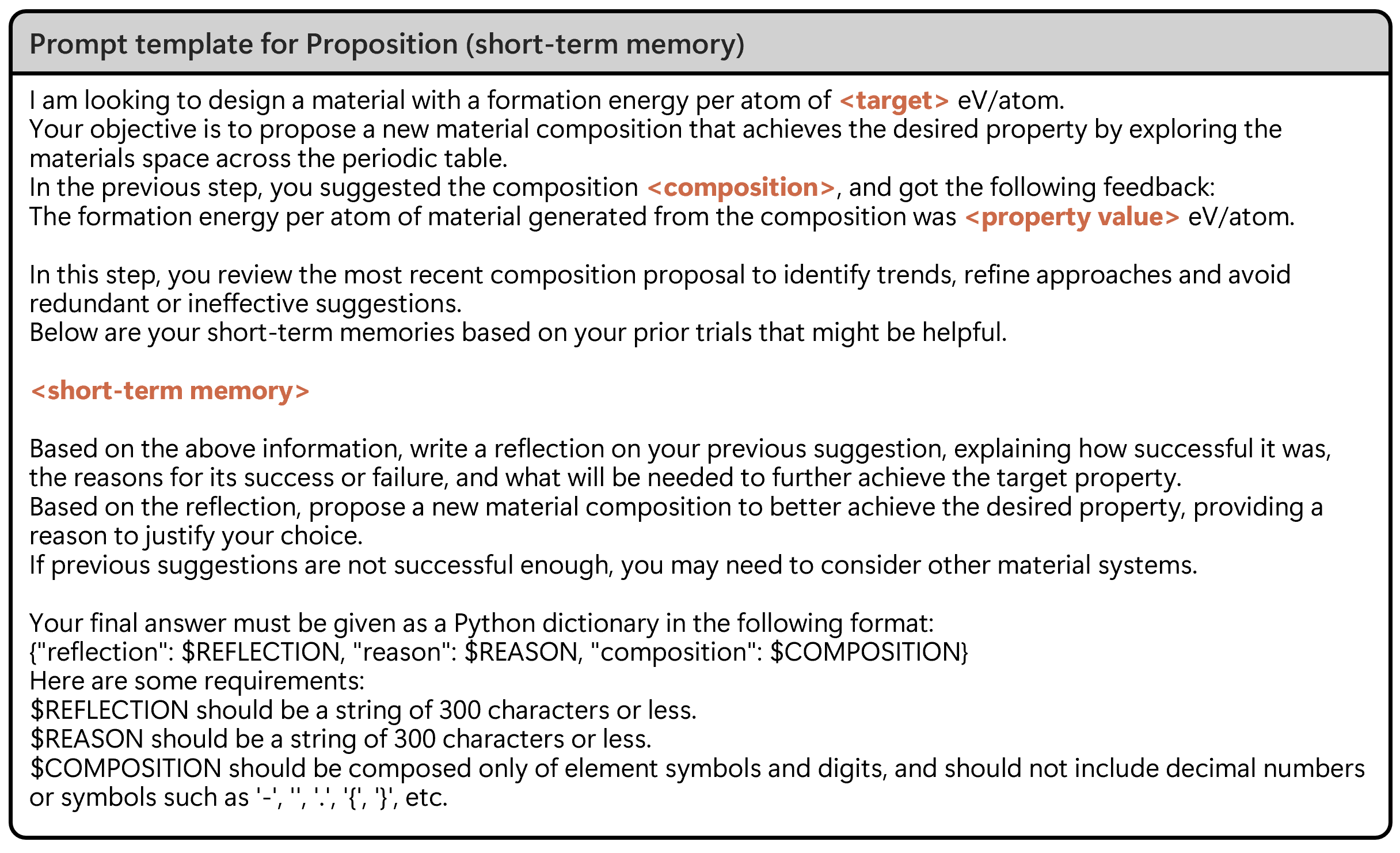}
    \caption{Prompt template for the Proposition stage when short-term memory has been selected. The template contains the following variables: \textbf{<target>}, \textbf{<composition>}, \textbf{<property value>}, and \textbf{<short-term memory>}. The \textbf{<short-term memory>} contains information regarding recent composition proposals, the reasoning behind those proposals, and the feedback received from the Property Evaluator.}
    \label{fig:prompt-sm}
\end{figure}

\begin{figure}[htbp]
    \centering
    \includegraphics[width=0.85\linewidth]{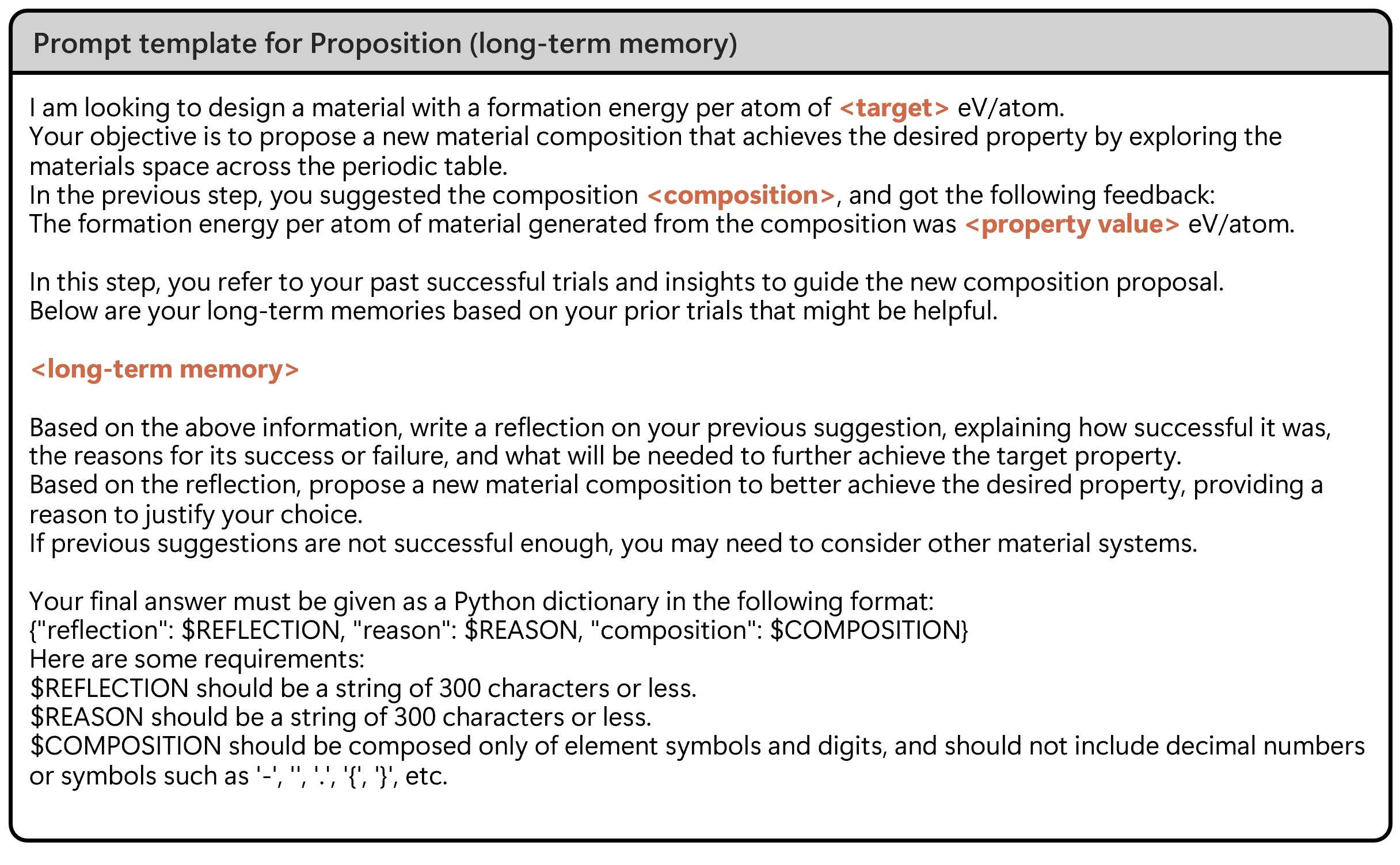}
    \caption{Prompt template for the Proposition stage when long-term memory has been selected. The template contains the following variables: \textbf{<target>}, \textbf{<composition>}, \textbf{<property value>}, and \textbf{<long-term memory>}. The \textbf{<long-term memory>} includes information on previous cases in which material compositions leading to structures with properties close to the target were successfully proposed, detailing the reasoning behind those proposals and the corresponding feedback received from the Property Evaluator.}
    \label{fig:prompt-lm}
\end{figure}

\begin{figure}[htbp]
    \centering
    \includegraphics[width=0.85\linewidth]{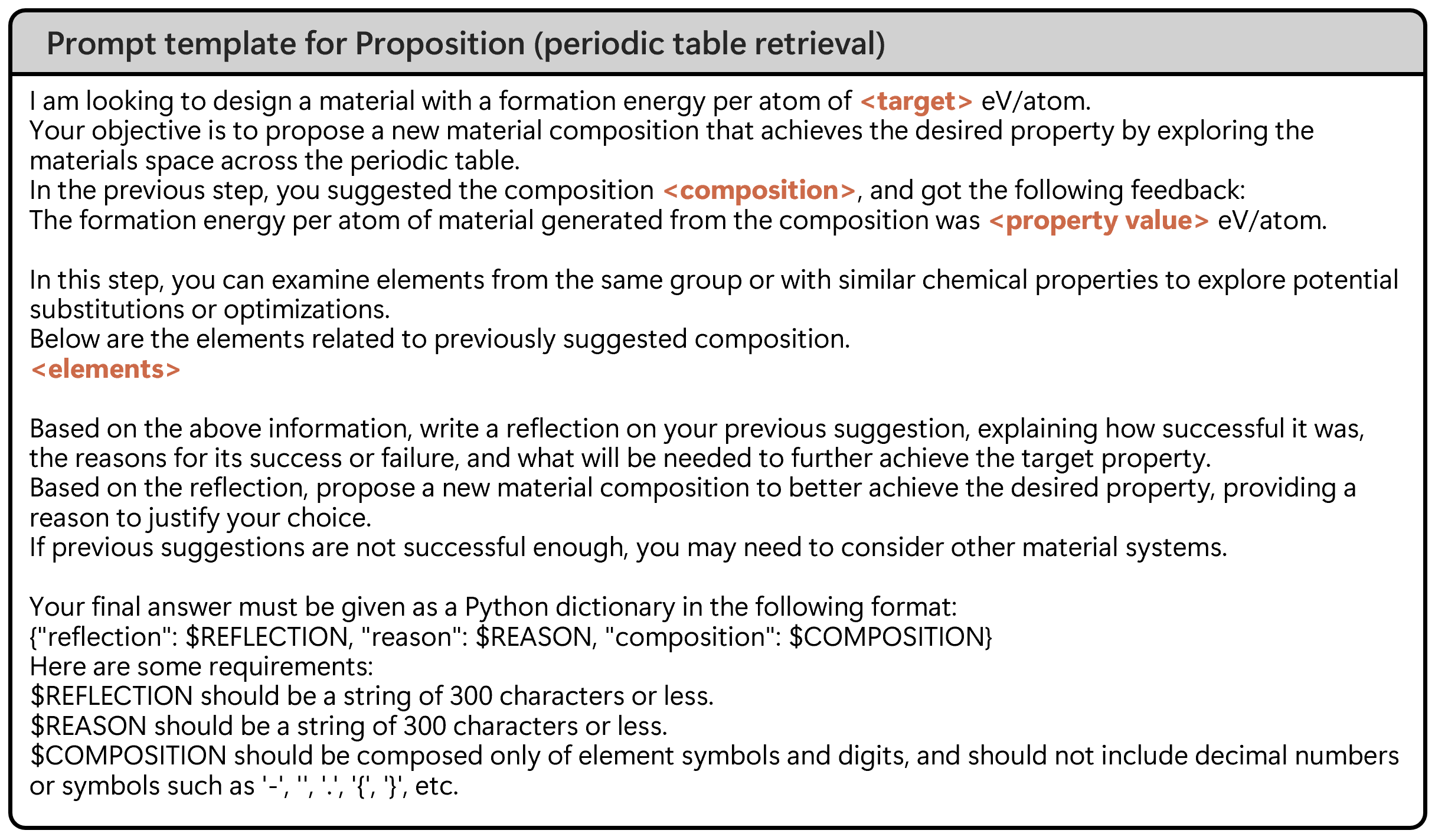}
    \caption{Prompt template for the Proposition stage when the periodic table has been selected. The template contains the following variables: \textbf{<target>}, \textbf{<composition>}, \textbf{<property value>}, and \textbf{<elements>}. The \textbf{<elements>} contains information about elements belonging to the same group as those included in the previous proposal. For example, if a previous proposal contains Li and O, \textbf{<elements>} will include Li: H, Na, K, Rb, Cs and O: S, Se, Te.}
    \label{fig:prompt-pt}
\end{figure}

\begin{figure}[htbp]
    \centering
    \includegraphics[width=0.85\linewidth]{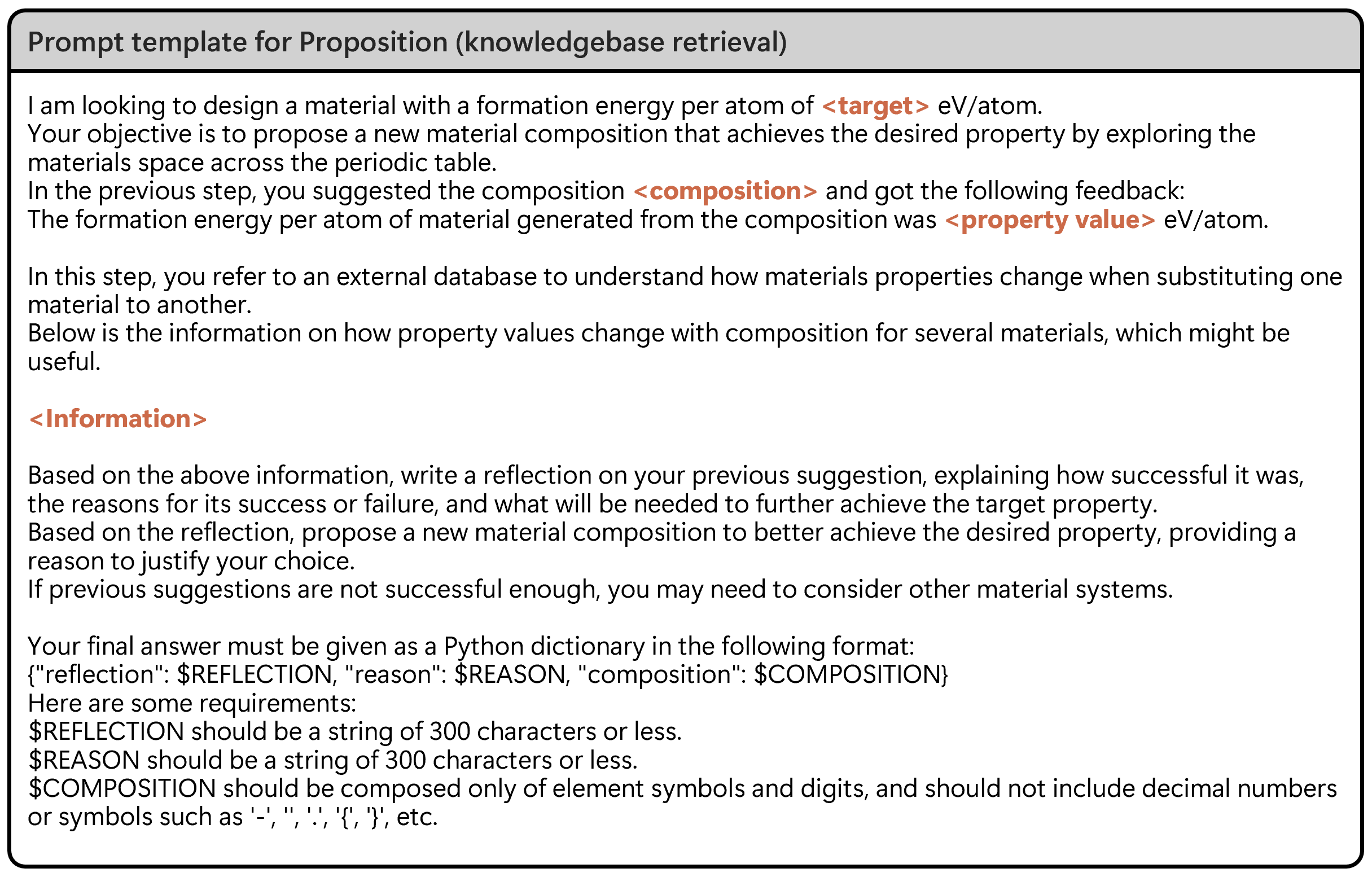}
    \caption{Prompt template for the Proposition stage when the materials knowledge base has been selected. The template contains the following variables: \textbf{<target>}, \textbf{<composition>}, \textbf{<property value>}, and \textbf{<Information>}. The \textbf{<Information>} includes insights into how material properties change when compositions related to the previously proposed composition are modified, along with observations explaining why these changes occur.}
    \label{fig:prompt-kb}
\end{figure}

Different prompt templates are used in the Proposition stage depending on which tool was selected during the Planning stage. The prompt template used when short-term memory is selected is shown in Figure \ref{fig:prompt-sm}. Similarly, the prompt templates used when long-term memory, periodic table, and materials knowledge base are selected are shown in Figures \ref{fig:prompt-lm}, \ref{fig:prompt-pt}, and \ref{fig:prompt-kb}, respectively.

\subsection{Prompt template for feedback generation}\label{sup:feedback}
\begin{figure}[htbp]
    \centering
    \includegraphics[width=0.85\linewidth]{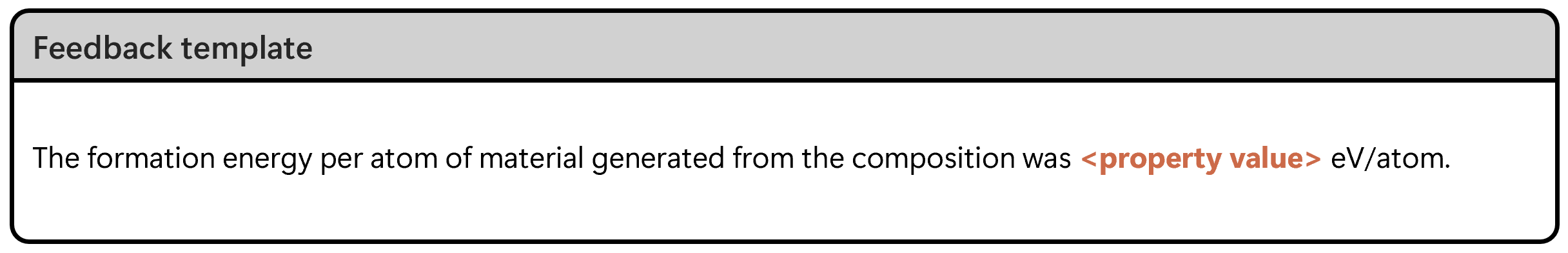}
    \caption{Template for feedback generation. The \textbf{<property value>} is a variable.}
    \label{fig:feedback}
\end{figure}

A prompt template used by the Property Evaluator to generate feedback for the LLM is shown in Figure \ref{fig:feedback}. This template is applied when the proposed composition is provided in a valid format. If the composition is submitted in an unreadable or incorrect format---for example, if it includes invalid element symbols or fractional subscripts---feedback instructing the LLM to resubmit the proposal using the correct format is provided.

\subsection{Prompt template for comparative experiments}\label{sup:baseline}
\begin{figure}[htbp]
    \centering
    \includegraphics[width=0.85\linewidth]{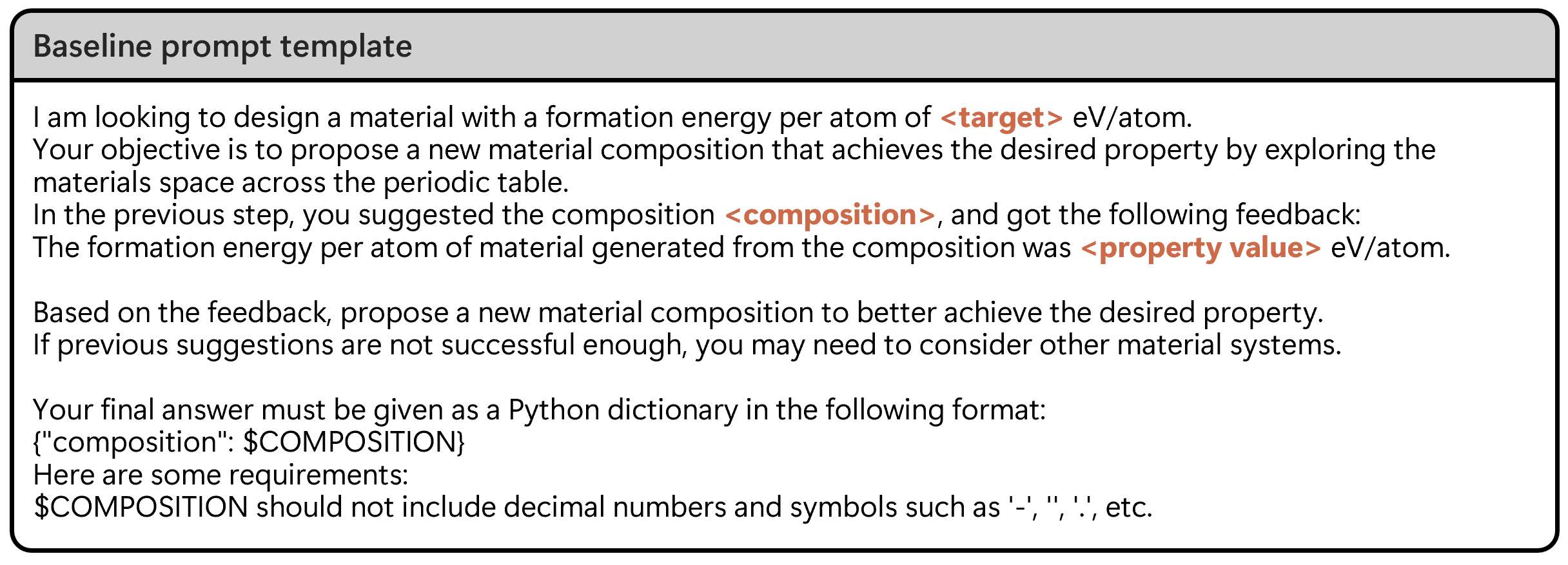}
    \caption{Prompt template for material composition proposal by the LLM without tool-assisted Planning and Proposition. \textbf{<target>}, \textbf{<composition>}, and \textbf{<property value>} are variables.}
    \label{fig:base_prompt}
\end{figure}

Figure \ref{fig:base_prompt} shows the prompt template used for having the LLM propose new compositions without tool-assisted Planning and Proposition. In this case, the LLM receives feedback on the previously proposed composition and, based on that feedback, returns a revised composition proposal.

\subsection{Prompt template for observation generation}\label{sup:transition}
\begin{figure}
    \centering
    \includegraphics[width=0.85\linewidth]{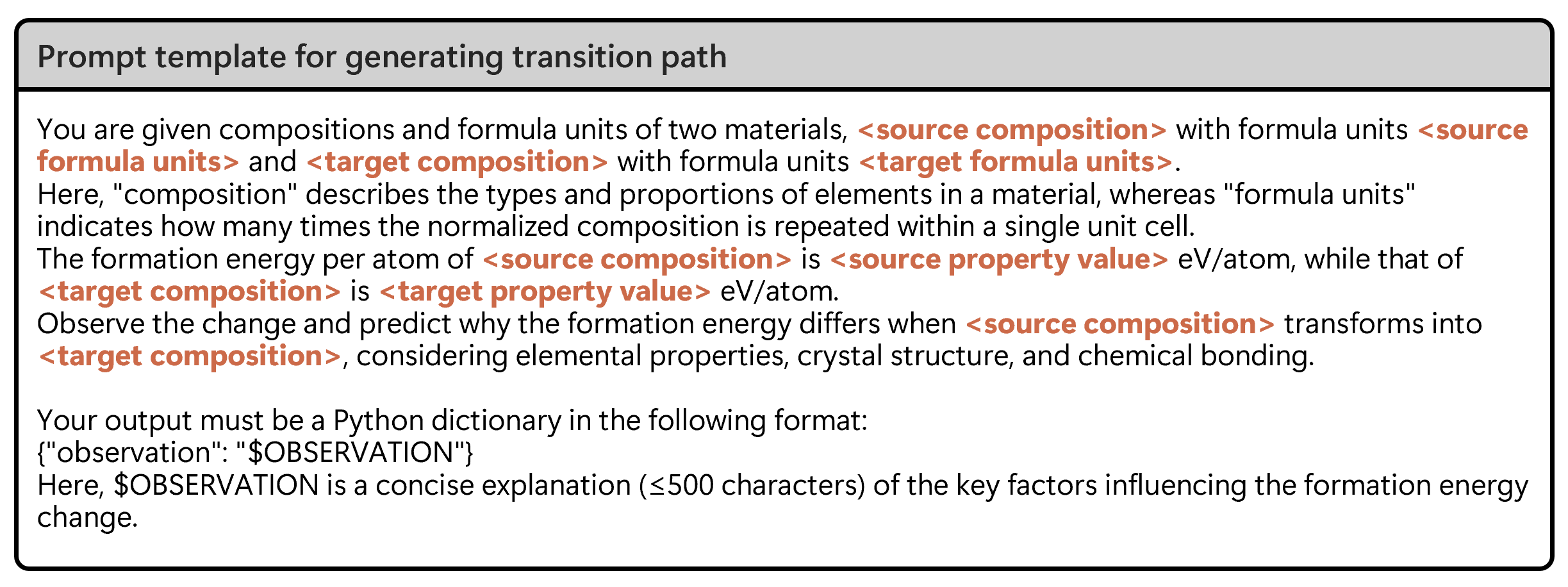}
    \caption{Prompt template used to generate explanations of observations when transitioning from one composition to another during the construction of the knowledge base. The variables are follows: \textbf{<source composition>}, \textbf{<source formula units>}, \textbf{<source property value>}, \textbf{<target composition>}, \textbf{<target formula units>}, and \textbf{<target property value>}.}
    \label{fig:transition}
\end{figure}

Figure \ref{fig:transition} shows the prompt template used for generating observation when one composition transitions to another during the construction of the knowledge base.

%% file: body/4_appendix_3.tex
\newpage
\section{Data distribution}\label{sup:data-dist}

\begin{figure}[htbp]
    \centering
    \includegraphics[width=0.5\linewidth]{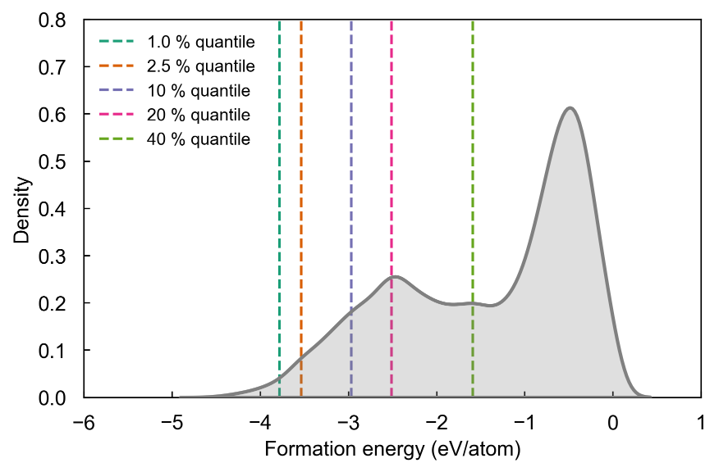}
    \caption{Distribution of formation energy per atom in MP-60 dataset.}
    \label{fig:eform-dist}
\end{figure}

Figure \ref{fig:eform-dist} shows the distribution of formation energies in the MP-60 dataset. The dashed lines indicate the values corresponding to the 1.0\% ($-$3.8 eV/atom), 2.5\% ($-$3.5 eV/atom), 10\% ($-$3.0 eV/atom), 20\% ($-$2.5 eV/atom), and 40\% ($-$1.6 eV/atom) quantiles, calculated by sorting the formation energy values in ascending order.

%% file: body/4_appendix_2.tex
\section{Reasoning examples}\label{sup:reason}

\begin{figure}[htbp]
    \centering
    \includegraphics[width=0.85\linewidth]{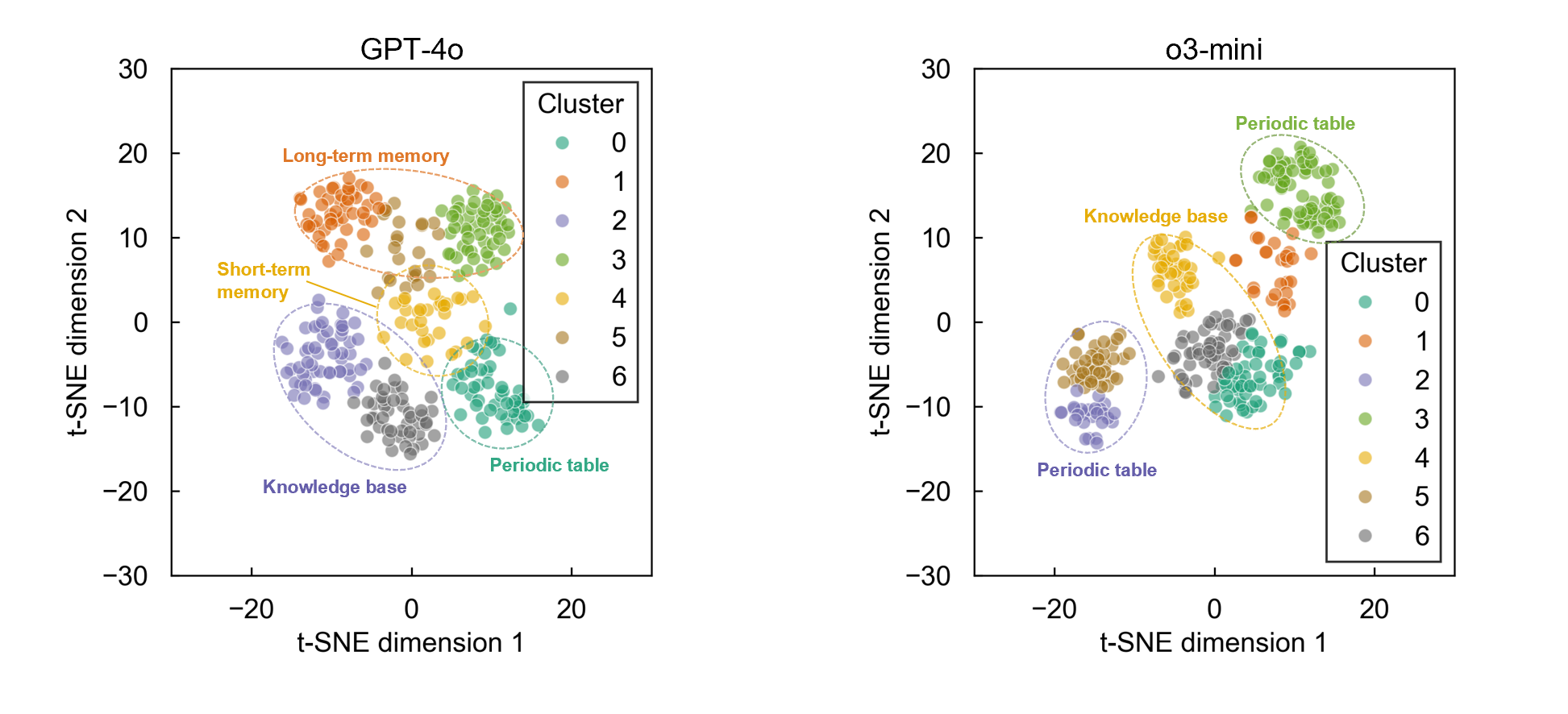}
    \caption{t-SNE visualization of the embedding representations of reasoning texts generated by the LLM during the Planning stage for GPT-4o and o3-mini. Each point represents a reason associated with a selected tool, and the colors indicate clusters obtained via k-means clustering. }
    \label{fig:tsne-plan}
\end{figure}

\begin{figure}[htbp]
    \centering
    \includegraphics[width=1.0\linewidth]{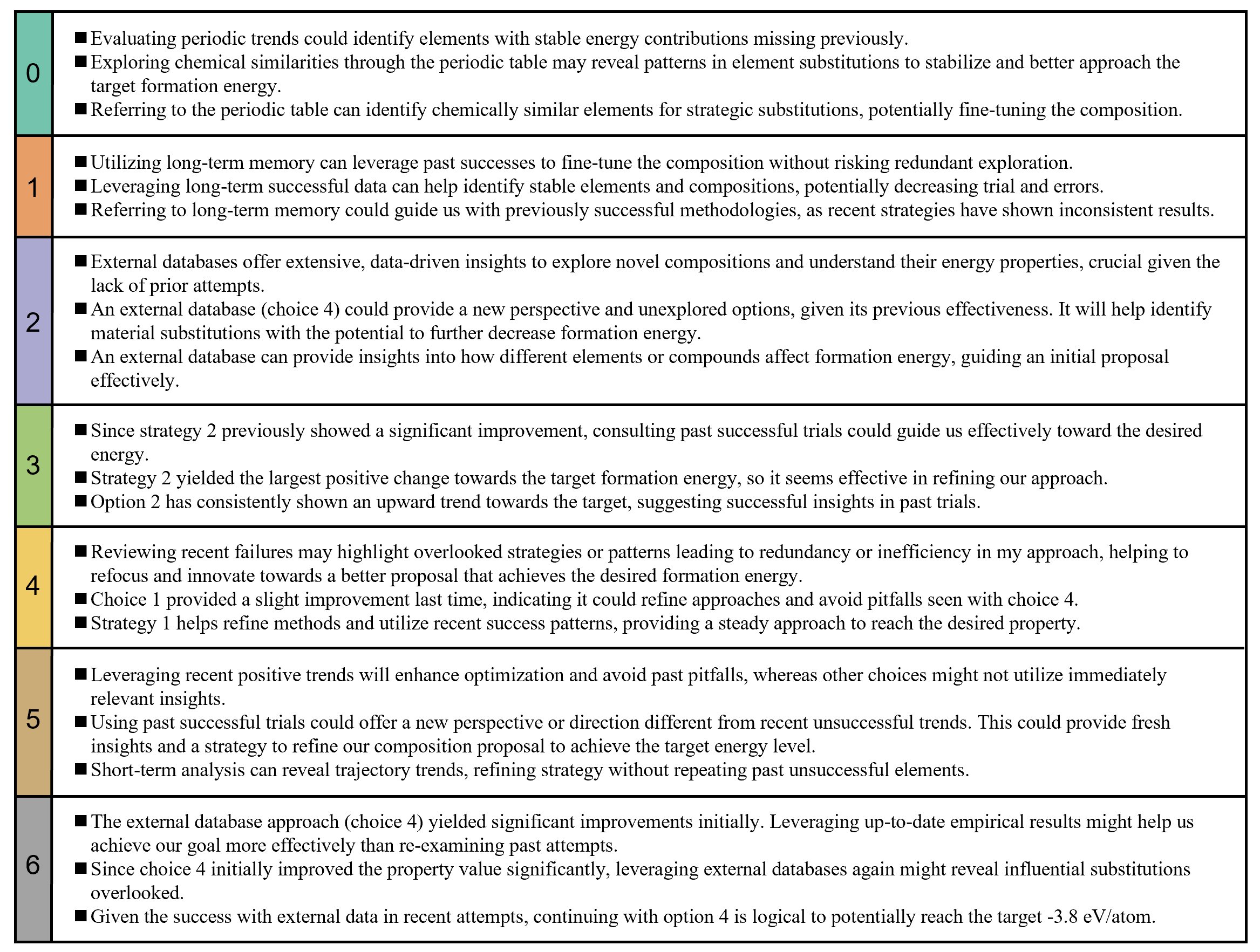}
    \caption{Examples of reasoning texts generated by GPT-4o during the Planning stage. For each cluster shown in Figure \ref{fig:tsne-plan}, three reasoning examples were randomly selected from the points belonging to that cluster. }
    \label{fig:reason-plan-gpt4o}
\end{figure}

\begin{figure}[htbp]
    \centering
    \includegraphics[width=1.0\linewidth]{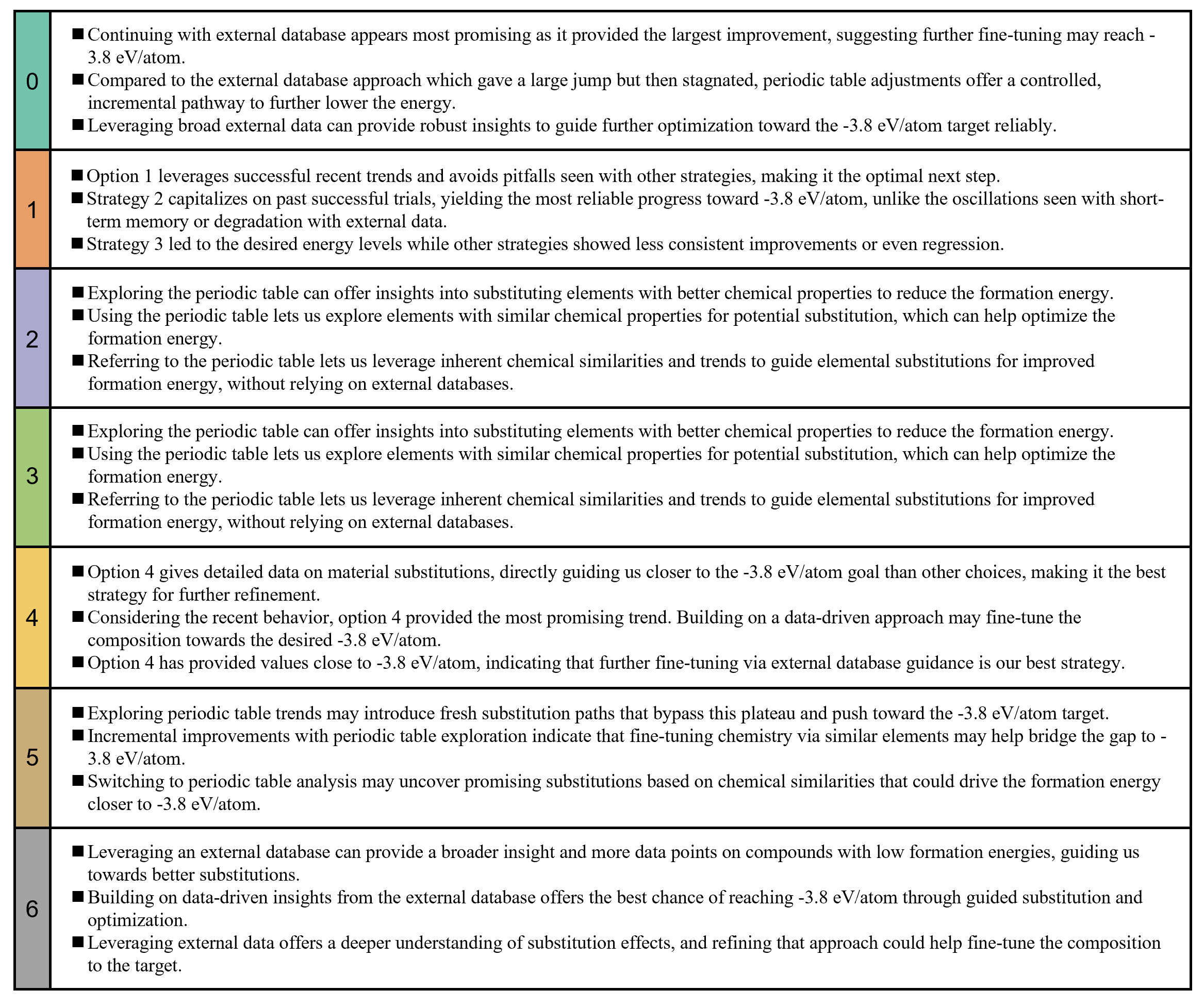}
    \caption{Examples of reasoning texts generated by o3-mini during the Planning stage. For each cluster shown in Figure \ref{fig:tsne-plan}, three reasoning examples were randomly selected from the points belonging to that cluster. }
    \label{fig:reason-plan-o3mini}
\end{figure}

One of the key advantages of the proposed framework lies in its interpretability, which stems from employing an LLM as the central reasoning engine. To investigate how the LLM reasons during the materials design process, we analyzed concrete examples of the reasoning outputs generated by the LLMs. Specifically, in the Planning stage, we collected the reasoning texts associated with each selected tool, converted them into embedding representations using SciBERT\cite{beltagy2019}, and applied t-SNE\cite{maaten2008} to project them into a two-dimensional space. The resulting visualization is shown in Figure \ref{fig:tsne-plan}. As can be confirmed from Figure \ref{fig:tsne-plan}, the embedding representations form distinct clusters, and the k-means clustering was applied to the embedded reasoning representations. The color of each point in the figure corresponds to its assigned cluster. To further interpret the characteristics of each cluster, raw reasoning texts associated with points in each cluster are shown in Figure \ref{fig:reason-plan-gpt4o} for GPT-4o and in Figure \ref{fig:reason-plan-o3mini} for o3-mini.

\begin{figure}[htbp]
    \centering
    \includegraphics[width=0.85\linewidth]{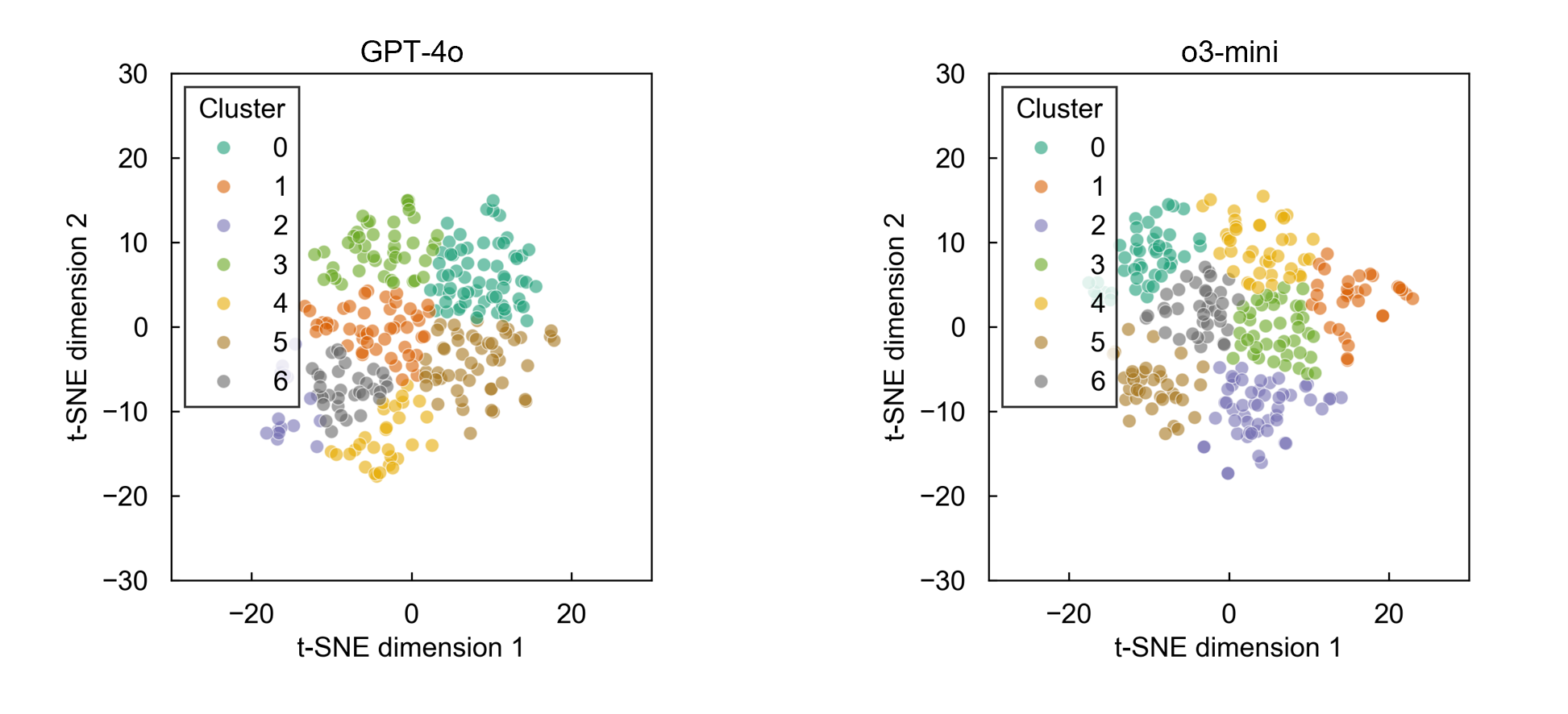}
    \caption{t-SNE visualization of the embedding representations of reasoning texts generated by the LLM during the Proposition stage for GPT-4o and o3-mini. Each point represents a reason associated with a selected tool, and the colors indicate clusters obtained via k-means clustering}
    \label{fig:tsne-propose}
\end{figure}

By examining the reasoning texts within each cluster in Figures \ref{fig:reason-plan-gpt4o} and \ref{fig:reason-plan-o3mini}, it becomes evident that each cluster corresponds to a specific tool selected by the LLM, indicating that the reasoning patterns are correlated with tool choice. Additionally, in Figure \ref{fig:reason-plan-gpt4o}, when the knowledge base is selected, the reasoning texts are further divided into two distinct clusters (purple Cluster 2 and gray Cluster 6). The texts in the gray cluster suggest that the LLM selected the knowledge base again based on prior experience, where using it previously led to successful outcomes. A similar trend was also observed for o3-mini as well. These results suggest that the LLM selects tools by taking past outcomes into account, demonstrating experience-informed decision-making.

\begin{figure}[htbp]
    \centering
    \includegraphics[width=1.0\linewidth]{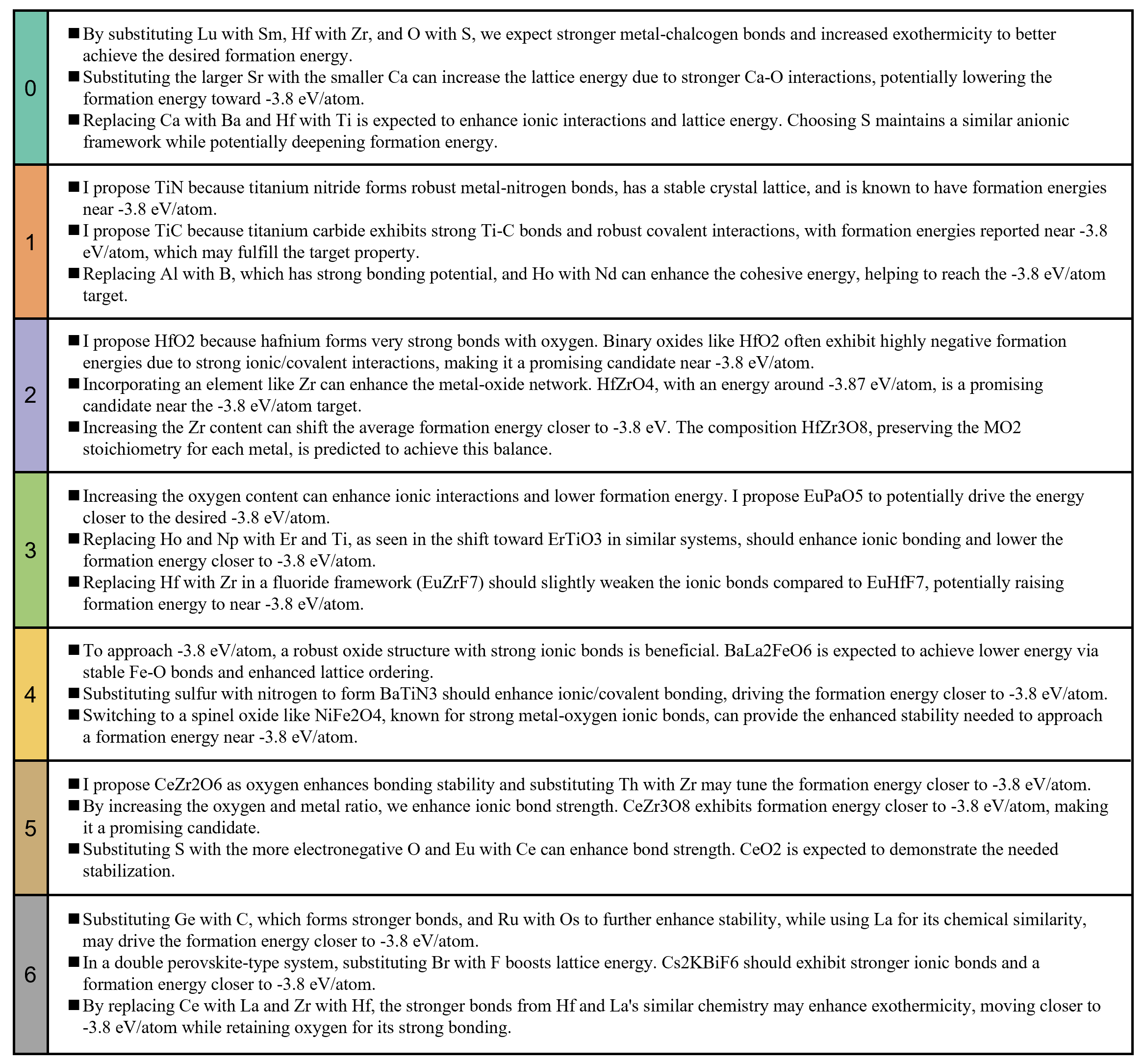}
    \caption{Examples of reasoning texts generated by GPT-4o during the Proposition stage. For each cluster shown in Figure \ref{fig:tsne-propose}, three reasoning examples were randomly selected from the points belonging to that cluster. }
    \label{fig:reason-propose-gpt4o}
\end{figure}

\begin{figure}[htbp]
    \centering
    \includegraphics[width=1.0\linewidth]{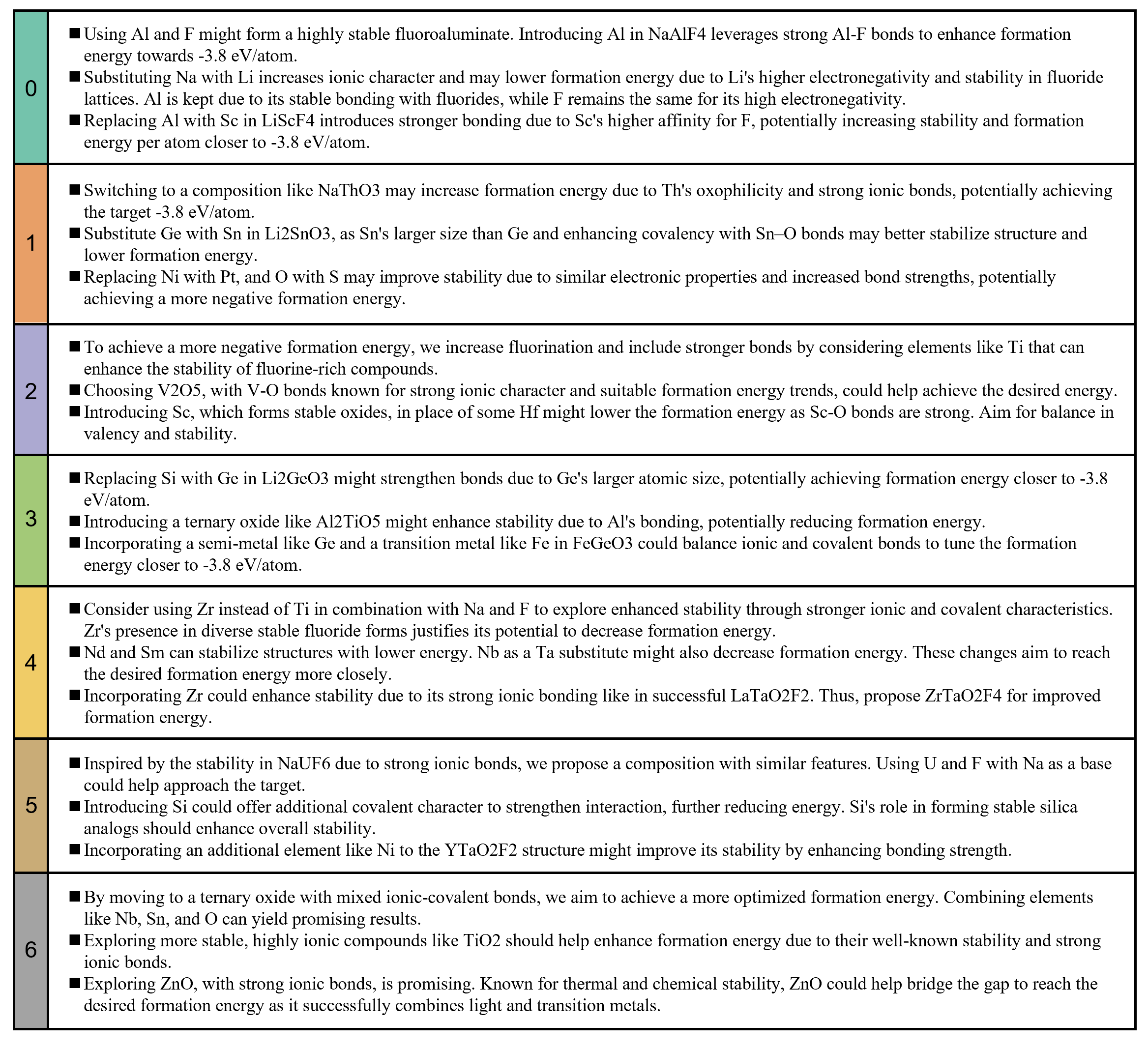}
    \caption{Examples of reasoning texts generated by o3-mini during the Proposition stage. For each cluster shown in Figure \ref{fig:tsne-propose}, three reasoning examples were randomly selected from the points belonging to that cluster. }
    \label{fig:reason-propose-o3mini}
\end{figure}

A similar analysis conducted for the Proposition stage is shown in Figure \ref{fig:tsne-propose}. Unlike in the Planning stage, the reasoning texts associated with composition proposals in the Proposition stage did not form clearly separatable clusters in the t-SNE analysis. This is likely because tool selection is a task involving a choice among four fixed options, whereas composition proposal is a more flexible task. For both GPT-4o and o3-mini, Figures \ref{fig:reason-propose-gpt4o} and \ref{fig:reason-propose-o3mini} show randomly selected reasoning texts from each region obtained by k-means clustering in the t-SNE space. Upon examining the reasoning texts, it was observed that texts containing words such as "substitution" or "replace" tend to be embedded in nearby regions. Although the physical and chemical correctness of the reasoning is not guaranteed in this study, it was observed that the LLM provides explanations using relevant keywords such as types of chemical bonding (e.g., ionic or covalent bonds), electronegativity, and atomic affinity. By further refining these explanations and improving their reliability, LLMs have the potential to provide valuable insights that human experts can interpret and effectively leverage in the materials design process.

